\documentclass[twocolumn,twoside,preprintnumbers,amsmath,amssymb,showkeys]{revtex4}
\usepackage{epsfig}
\usepackage{graphicx}
\usepackage{dcolumn}
\usepackage{bm}

\usepackage{fancyhdr}

\usepackage{pslatex}

\begin{document}

\title{{ QCD Green Functions and their Application to Hadron Physics }}

\author{Reinhard Alkofer}

\affiliation{Institut f\"ur Physik, Universit\"at Graz, 
Universit\"atsplatz 5,
A-8010 Graz, Austria }

\received{on 7 November, 2006}

\begin{abstract}

In a functional approach to  QCD the infrared behaviour of Landau gauge Green
functions is investigated. It can be proven that the ghost Dyson-Schwinger
equation implies the Gribov-Zwanziger horizon condition. Its relation to the
Kugo-Ojima confinement scenario is elucidated. Positivity violation for gluons 
is demonstrated, and the analytic structure of the gluon propagator is studied.
Quark confinement is related to an infrared divergence of the quark-gluon
vertex. It is shown that in the latter various components are non-vanishing due
to the dynamical breaking of chiral symmetry.
As a result an infrared finite running coupling in the Yang-Mills sector
is derived whereas the running coupling related to the quark-gluon vertex is
infrared divergent.
In Coulomb gauge QCD already the one-gluon-exchange (over-)confines.
This leads to a vanishing quark propagator, and thus quarks are confined.
Nevertheless colour singlet quantities derived from the quark propagator
are well-defined. Especially the expression for the quark condensate proves
that chiral symmetry is dynamically  broken. As expected the properties of
mesons can be directly calculated whereas the mass of
coloured diquarks diverges, and thus diquarks are confined. The latter 
nevertheless possess a well-defined size.
In the third part the results obtained so far will be  used to
formulate a covariant Faddeev approach to nucleons. The resulting amplitudes
describe the quark core of the nucleon. Besides the mass of this state also
the electromagnetic form factors are calculated. The results for charge radii 
and magnetic moments as a function of the quark current mass provide some
indication what the missing pion cloud may contribute to the nucleons'
properties.




\end{abstract}

\maketitle

\thispagestyle{fancy}
\setcounter{page}{0}

\section{On Confinement in the Covariant Gauge:
Analytic properties of the Landau gauge gluon and quark propagators}

\subsection{Motivation}
Hadrons are believed to consist of quarks and gluons. However, every attempt to
break a hadron into its constituents has failed so far. Thus the situation is
very much different as {\it e.g.\/} in atomic physics. There studies of
seperated atomic nuclei and electrons directly verify the nature of atoms being
bound states. In hadronic physics, however, the evidence of physical states
being bound states is indirect. Especially the plethora of hadrons
finds a natural explanation in the assumption of hadrons being composite but
their constituents do not exist as free particles. This phenomenon is
called {\em confinement}.

The focus of these lectures is the infrared behaviour of QCD~Green functions
(for recent reviews see {\it e.g.\/} refs.\
\cite{Alkofer:2000wg,Fischer:2006ub}) and what they tell us about confinement.
It will be seen that from this perspective the confinement mechanisms for
gluons, quarks, and colored composites show some distinctive features. Another
virtue of the presented approach is the fact that QCD Green functions can be
employed directly in hadron phenomenology. This opens up a road to the
ambitious aim of verifying or falsifying confinement pictures experimentally.

A property of QCD which can also be infered from the hadron spectrum is the
dynamical breaking of chiral symmetry. A correct description of this
non-perturbative phenomenon within functional approaches requires that the 
chiral Ward identities between Green functions are respected, see {\it e.g.\/}
refs.\  \cite{Alkofer:2000wg,Maris:2003vk}. Schemes
fulfilling this condition exist, and the resulting description of meson physics
is impressingly successful (although many properties of mesons still have to be
understood). Recently the possibility of an ab initio description of the
nucleon in functional approaches to continuum quantum field theory has opened
up, and my third lecture will describe the first few steps in such a
direction. 

The confinement problem has proven to be notoriously difficult, see {\it
e.g.\/} ref.\ \cite{Alkofer:2006fu} and references therein for a brief  review of
(some of) the currently investigated theories of confinement.  To mention two
reasons out of many why the confinement problem is especially hard to solve 
let me first remark that the length scale of confinement is a physical scale.
Based on renormalization group (RG) considerations one can then conclude that
there has to exist an RG invariant confinement scale which is related to the
renormalization scale $\mu$ via the $\beta$-function \cite{Gross:1974jv}:
\begin{equation} 
\Lambda_{\tt conf} = \mu \exp \left( - \int ^g \frac
{dg'}{\beta (g')} \right) \stackrel{g\to 0}{\rightarrow } \mu  \exp \left( -
\frac 1 {2\beta_0g^2} \right) \ .
\end{equation} 
This relation clearly shows that
the confinement scale possesses an essential  singularity in the coupling
constant $g$. Therefore it is not accessible in perturbation theory, and
confinement cannot be described perturbatively.

In addition, as anticipated and also verified in the course of these lectures, 
confinement is related to infrared singularities. Numerical lattice Monte-Carlo
calculations are always restricted to finite volumes, and infrared properties
can only be investigated by carefully studying the infinite volume limit. From
this remark it is obvious that, to complement the lattice approach, a  {\em
non-perturbative continuum} approach is highly desirable if one aims at an 
understanding of confinement.

\subsection{Basic Concepts}

\subsubsection{Covariant Gauge Theory}
To obtain confinement the least requirement on the fundamental fields of QCD
is that they do not represent particles, or phrased otherwise, that they do not
appear as asymptotic states in the $S$-matrix.

\begin{figure}[htbp]
\begin{center}
\includegraphics[width=8cm]{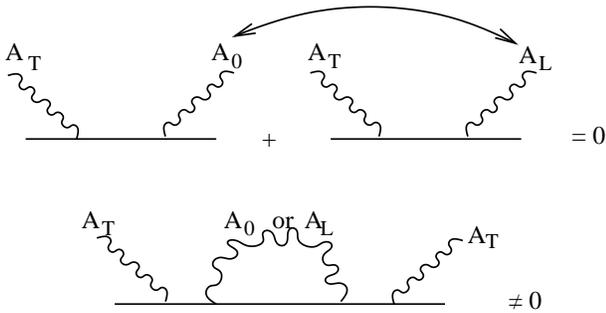}
\caption{An example of Feynman diagrams containing longitudinal and timelike
photons as external lines (where they cancel) and in loops (where they 
contribute).}
\label{photons}
\end{center}
\end{figure}

As a matter of fact, some relations between different confinement scenarios 
become most transparent in a
covariant formulation which includes the choice of a covariant gauge, of
course. First we note that covariant quantum theories of gauge fields require
indefinite metric spaces. Abandoning the positivity of the representation space
already implies to give up one of the axioms of standard quantum field theory.
Maintaining the much stronger principle of locality, gluon confinement then
naturally relates to the violation of positivity in the gauge field sector, see
{\it e.g.\/} ref.\ \cite{Alkofer:2000wg}. A comparison to QED, where the
Gupta-Bleuler prescription \cite{Bleuler50} is to enforce the Lorentz condition
on physical states, is instructive. There
a semi-definite physical subspace can be defined as the
kernel of the (onto its positive frequency part projected)
field operator $\partial _\mu A^\mu$.
The physical states $|\Psi \rangle $ fulfilling
\begin{equation}
\partial _\mu A^\mu |\Psi \rangle = 0
\end{equation}
then correspond to (equivalence
classes of) states in this subspace. Covariance implies, besides transverse
photons, the existence of longitudinal and timelike (``scalar'') photons in
QED. The latter two form metric partners in the indefinite space: They cancel
against each other in every $S$-matrix element and therefore do not contribute
to observables. To be more precise: The unphysical states have to be kept
when inserting a complete unity (in the language of Feynman diagrams, in loops).
They destructively interfere in  between amplitudes ({\it i.e.\/} Feynman
diagrams) containing  these states as asymptotic states ({\it i.e.\/} 
external lines). A simple example of such Feynman diagrams is given in fig.\
\ref{photons}.

In QCD cancelations of unphysical degrees of freedom in the $S$-matrix  also
occur but are more complicated due to the self-interaction of the gluons,
transverse gluons scatter into longitudinal ones and vice versa. In
perturbation theory one obtains {\it e.g.\/} amplitudes for the scattering  of
two transverse into one transverse and one longitudinal gluons to order
$\alpha_S^2$. A consistent quantum formulation in a functional integral
approach leads to the introduction of ghost fields $c^a$ and $\bar c^b$
\cite{Faddeev67}. To order
$\alpha_S^2$ a ghost loop then cancels all gluon loops which describe
scattering of transverse to longitudinal gluons. The proof of this cancelation
to all orders in perturbation theory has been possible by employing the BRST
symmetry of the covariantly gauge fixed theory~\cite{Becchi:1976nq}. At this
point one has achieved a consistent quantization.  Also one should note that
renormalizibility rests on BRST symmetry. 

It is useful to picture the BRST transformation $\delta_B$
as a ``gauge transformation''
with a constant ghost field as parameter
\begin{equation}
\begin{array}{cc}
\delta_B A^a_\mu \, =\,  D^{ab}_\mu c^b \, \lambda  \; , \quad &  \delta_B q
\, = \,-  i g t^a \, c^a \, q \, \lambda \; , \\
\delta_B c^a \, = \, - \, \frac{g}{2} f^{abc} \, c^b c^c \, \, \lambda \; ,
\quad  & \delta_B\bar c^a \, = \, \frac{1}{\xi} \partial_\mu A_\mu^a
\, \lambda \; , \end{array}  
\end{equation}
where $D^{ab}_\mu$ is the covariant derivative and $\xi$ is the gauge-fixing 
parameter of linear covariant gauges.
The parameter $\lambda$ lives in the Grassmann algebra of the ghost fields, 
it carries ghost number $N_{\mbox{\tiny FP}} = -1$. Via the Noether theorem
one may define a BRST charge operator $Q_B$ which in turn generates a ghost
number graded algebra on the fields, $\delta_B\Phi = \{ i Q_B, \Phi \}$.
Defining the ghost number operator $Q_c$ one obtains
\begin{equation}
           Q_B^2 = 0 \; , \quad \left[ iQ_c , Q_B \right] = Q_B \; .
\end{equation}
This algebra is complete in the indefinite metric state space $\mathcal V$.

As the gauge fixing Lagrangian is BRST exact,
\begin{equation}
{\mathcal L}_{GF} = \delta_B \left( \bar c \left( \partial_\mu A^\mu +
\frac \xi 2 B \right) \right) ,
\end{equation}
the proof of BRST invariance of the gauge fixed action is straightforward. 

The semi-definite  physical subspace  ${\mathcal{V}}_{\mbox{\tiny phys}}  =
\mbox{Ker}\, Q_B  $ is defined on the basis of this algebra by those states
which are annihilated by the BRST charge $Q_B$, $Q_B |\psi \rangle =0$. Since
$Q_B^2 =0 $, this subspace contains the space $ \mbox{Im}\, Q_B $ of so-called
daughter states $Q_B |\phi \rangle$ which are images of their parent states in
$\mathcal{V}$. A physical Hilbert space is then obtained as the space of
equivalence classes, the BRST cohomology of states in the kernel modulo those
in the image of $Q_B$,
\begin{equation}
     {\mathcal{H}}(Q_B,{\mathcal{V}}) = {\mbox{Ker}\, Q_B}/{\mbox{Im}\, Q_B}
       \simeq  {\mathcal{V}}_s \; .
\end{equation}
This Hilbert space is isomorphic to the space of BRST singlets. All states are
either BRST singlets or belong to quartets, this exhausts all possibilities.
This generalization of the Gupta-Bleuler condition on physical states, 
{\it i.e.\/} $Q_B |\psi\rangle = 0$, 
eliminates half of these metric partners from  all $S$-matrix
elements (leaving unpaired
states of zero norm which do not contribute to any observable). 

Are the transverse gluons also part of a BRST quartet? Are gluons confined this
way as conjectured in refs.\ \cite{Kugo:1979gm,Nakanishi:qm}? Before we return
to this question it is illustrative to have a closer look into the issue of gauge
fixing.

\begin{figure*}[htbp]
\begin{center}
\includegraphics[width=120mm,height=40mm]{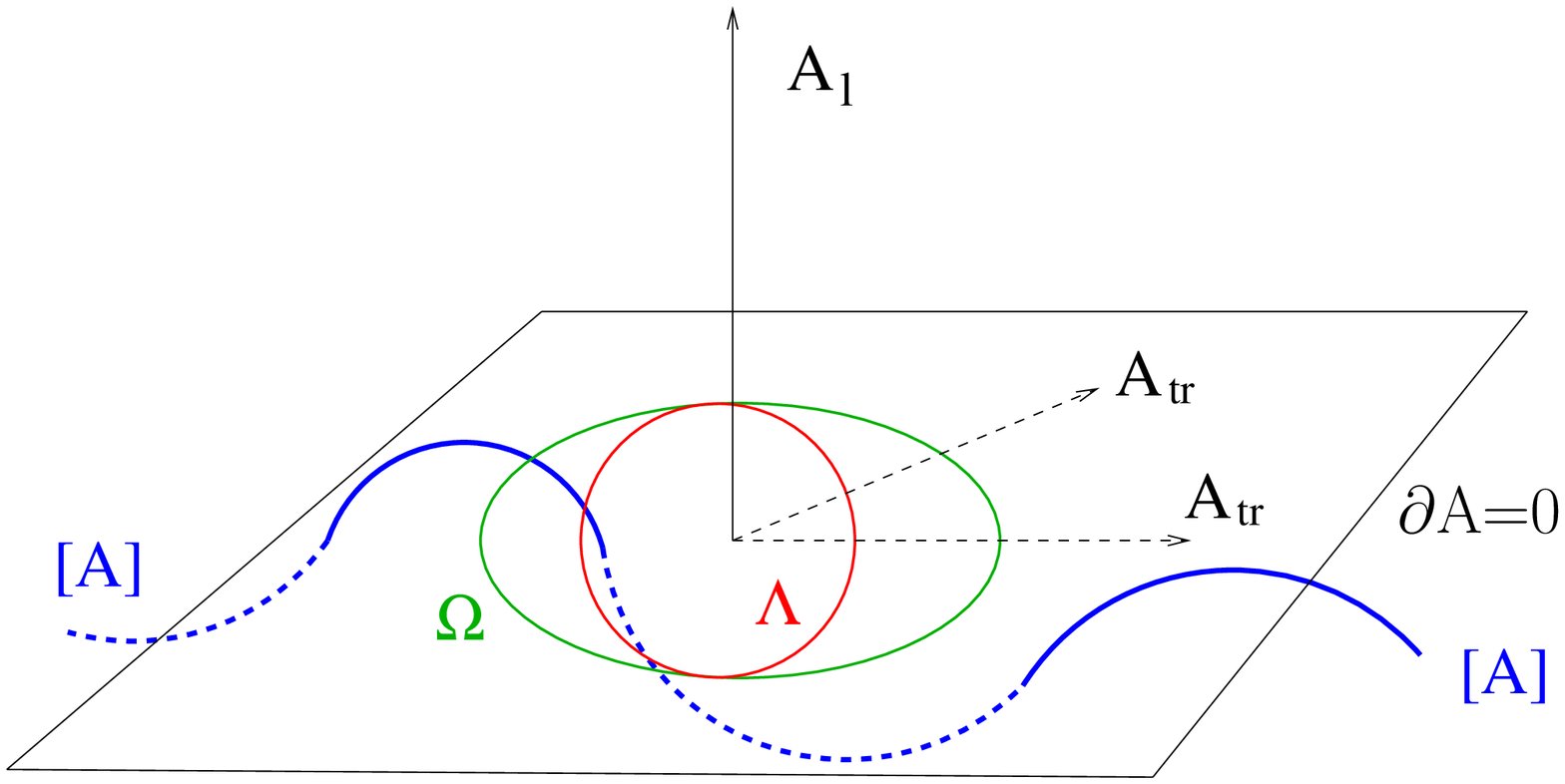}
\label{Gribov}
\caption{A schematic representation of the configuration space of gauge fields,
the ``hyperplane'' of transverse gauge fields, the first Gribov region and the
fundamental modular region.}
\end{center}
\end{figure*}

\subsubsection{Gribov horizon {\&} Zwanziger condition}
A detailed illustration of the issue of the non-uniqueness of gauge fixing
\cite{Gribov:1978wm}  has been given in the lectures by Dan Zwanziger
\cite{Dan}. Therefore the presentation given here focuses on the implications of
the infrared behaviour of QCD Green functions.
Within the state of all gauge field configurations the ones fulfilling the
na\"ive Landau gauge, {\it i.e.\/} the transverse gauge fields, form a
``hyperplane'' $\Gamma = \{A:\partial \cdot A=0\}$. As schematically
illustrated in fig.~2
a gauge orbit intersects $\Gamma$ several
times and therefore gauge fixing is not unique. The so-called minimal Landau
gauge obtained by minimizing $||A||^2$ along the gauge orbit is usually 
employed in corresponding lattice
calculations, it restricts the gauge fields to the Gribov region
\begin{equation}
\Omega = \{A: ||A||^2 \, minimal\} = \{A: \partial \cdot A=0,
-\partial \cdot D (A) \ge 0 \}
\end{equation}
where the Faddeev operator $-\partial \cdot D (A)$ is strictly positive definite.
Phrased otherwise: On the boundary of the Gribov region, the Gribov horizon,
the Faddeev operator possesses at least one zero mode. 

Unfortunately this is not the whole story. There are still Gribov copies
contained in $\Omega$ \cite{vanBaal:1997gu,Williams:2002dw}, 
therefore one needs to restrict the gauge field
configuration space even further to the region of global minima of $||A||^2$
which is called the fundamental modular region. A lot of effort has been
devoted to an estimate of the corresponding effects in lattice studies,
see {\it e.g.\/}
\cite{Cucchieri:1997dx,Bonnet:2001uh,Alexandrou:2002gs,Silva:2004bv,Furui:2004cx,Bowman:2004jm,Sternbeck:2005tk}.
These provided evidence that for Green functions of lattice gauge theory the 
effects due to Gribov copies are not too large. The   situation may be even 
better for the continuum field theory: From an approach
using stochastic quantisation Zwanziger argued that Gribov copies inside the 
Gribov region have no effect on the Green functions of the theory 
\cite{Zwanziger:2003cf}. 

What remains to be implemented within a functional approach in continuum field
theory is to cut off the functional integral over gauge fields at the 
boundary  $\partial \Omega$ as already suggested by Gribov. The solution has
been found by Zwanziger \cite{Zwanziger:1993qr}: 
One has to require that the ghost propagator is more singular in the
infrared than a simple pole,
\begin{equation}
\lim_{k^2\to 0} \left(
k^2 D_{\text{Ghost}}(k^2) \right)^{-1} = 0.
\end{equation}

\subsubsection{Kugo--Ojima confinement criterion}

In the covariant gauge confinement depends on the realization of the unfixed
global gauge symmetries. The identification of the BRST singlets with color
singlets is possible only if the charge of global gauge transformations is BRST
exact {\em and} unbroken, {\it i.e.}, well-defined in the whole of the
indefinite metric space $\mathcal{V}$.  Then BRST singlets are the physical
states and are thus constituting the physical Hilbert space $\mathcal{H}$.
The sufficient conditions for this are provided by the Kugo-Ojima
criterion \cite{Kugo:1979gm,Nakanishi:qm}:  The current 
\begin{equation}
    J^a_\mu = \partial_\nu F_{\mu\nu}^a  + \{ Q_B , D_{\mu}^{ab} \bar c^b \}
     \; ,
       \label{globG}
\end{equation}
is globally conserved, $\partial_\mu J^a_\mu = 0 $, and its two terms are of 
the form of a total derivative. The first term corresponds to a coboundary
with respect to the space-time exterior derivative while the second term
is a BRST coboundary. The corresponding charges are 
denoted by $G^a$ and $N^a$, respectively,
\begin{equation}
      Q^a =  \int d^3x \,  \partial_i F_{0 i}^a \,  +\,  \int d^3x \,
             \{ Q_B , D_{0}^{ab} \bar c^b \} \, = \, G^a \, + \, N^a \; .
        \label{globC}
\end{equation}
For the first term herein there are only two possibilities, 
it is either ill-defined due to massless states in the spectrum of the 
field operator $\partial_\nu F_{\mu\nu}^a $, or else
it vanishes.

In QED massless photon states contribute to the analogues of both currents
in~(\ref{globG}), and both charges on the r.h.s.\ in (\ref{globC}) are
separately ill-defined. One can employ an arbitrariness in the definition of
the generator of the global gauge transformations (\ref{globC}) to multiply the
first term by a suitable constant such that these massless contributions
cancel. This way one obtains a well-defined and unbroken global gauge charge
which replaces the na\"ive definition in (\ref{globC}).  There are two
independent structures in the globally conserved gauge currents in QED which
both contain massless photon contributions. These can be combined to yield one
well-defined charge as the generator of global gauge transformations leaving
any other combination spontaneously broken, such as the displacement symmetry
which leads to the identification of the photon with massless Goldstone bosons
\cite{Nakanishi:qm}.

In case the term $\partial_\nu F_{\mu\nu}^a $ contains no massless
discrete spectrum, {\it i.e.}, if there is no massless
particle pole in the Fourier transform of transverse gluon correlations, 
one obtains $G^a \equiv 0$.
In particular, this is the case for channels containing massive vector fields,
{\it i.e.\/} in theories with Higgs mechanism. It is expected to be also the 
case in
any color channel of QCD with confinement for which it actually represents one
of the two conditions formulated by Kugo and Ojima.
In both these situations the BRST exact charge is
\begin{equation}
                       Q^a \, = \, N^a \, = \, \Big\{   Q_B \, ,
       \int d^3x \,   D_{0}^{ab} \bar c^b \Big\} \; .
\end{equation}
The second of the two conditions for confinement is that this charge be
well-defined in the whole of the indefinite metric space $\mathcal{V}$.
Together these conditions are sufficient to establish that all BRST singlet
physical states in $\mathcal{H}$ are also colour singlets, and that all coloured
states are thus subject to the quartet mechanism. The second condition thereby
provides the essential difference between Higgs mechanism and confinement. The
operator $D_\mu^{ab}\bar c^b$ determining the charge $N^a$ will in general
contain a  {\em massless} contribution from the elementary quartet due to the
asymptotic field $\bar\gamma^a(x)$ in the antighost field which is defined 
in the (weak)  asymptotic limit: $\bar c^a\,
\stackrel{x_0 \to \pm\infty}{\longrightarrow} \, \bar\gamma^a + \cdots $.
In the resulting relation 
\begin{equation}
          D_\mu^{ab}\bar c^b \; \stackrel{x_0 \to \pm\infty}{\longrightarrow}
              \;   ( \delta^{ab} + u^{ab} )\,   \partial_\mu \bar\gamma^b(x) +
                 \cdots  \;  
\end{equation}
the dynamical parameters $ u^{ab} $ determine the contribution
of the massless asymptotic state to the composite field operator 
$g f^{abc} A^c_\mu
\bar c^b  \, \stackrel{x_0 \to \pm\infty}{\longrightarrow}  \,
u^{ab} \partial_\mu \bar\gamma^b + \cdots $. These parameters can be obtained
in the limit $p^2\to 0$ from the Euclidean correlation functions of this
composite field, {\it e.g.},
\begin{widetext}
\begin{equation}
\int d^4x \; e^{ip(x-y)} \,
\langle  \; D^{ae}_\mu c^e(x) \; gf^{bcd}A_\nu^d(y) \bar c^c (y) \; \rangle
\; =: \; 
\Big(\delta_{\mu \nu} - \frac{p_\mu p_\nu}{p^2} \Big) \, u^{ab}(p^2)
\; .  \label{Corru}
\end{equation}
\end{widetext}
The theorem by Kugo and Ojima asserts that all $Q^a = N^a$ are
well-defined in the whole of  $\mathcal{V}$ (and do not suffer from
spontaneous breakdown), if and only if
\begin{eqnarray}
                 u^{ab} \equiv u^{ab}(0)  \stackrel{!}{=} - \delta^{ab} \; .
\label{KO1}
\end{eqnarray}
Then the massless states from the elementary quartet do not contribute to
the spectrum of the current in $N^a$, and the equivalence between physical
BRST singlet states and color singlets is established.

Within the described mechanism the  physical state space of Yang-Mills theory
contains only colourless states. The coloured states are not BRST singlets and
therefore do not appear in $S$-matrix elements, they are unobservable.  In the
following I will provide evidence that the transverse gluons are BRST quartet
states with gluon-ghost, gluon-antighost and gluon-ghost-antighost states in
the same multiplet. Gluon confinement then occurs as kind of destructive
interference between amplitudes ({\it i.e.\/} Feynman diagrams) containing 
these states as asymptotic states ({\it i.e.\/} external lines). The members of
quartets are frequently said to be ``confined kinematically''.
This BRST quartet mechanism can be summarized as follows:
\begin{itemize}
\item
Perturbatively, just as in QED, one such quartet, the elementary quartet, 
is formed by
the massless asymptotic states of longitudinal and timelike gluons together
with ghosts and antighosts which are thus all unobservable.
\item
Non-perturbatively, and in contrast to QED, however, the quartet mechanism 
also applies to
transverse gauge field, {\it i.e.\/} gluon,  states. A violation of 
positivity (see the next subsection) for such states
then entails that they are also unobservable.
\end{itemize}

In Landau gauge a sufficient criterion for relation (\ref{KO1}) and thus  for
this type of confinement to occur is given by the infrared behaviour of the
ghost propagator: If it is more singular than a simple pole the Kugo-Ojima
confinement criterion is fulfilled \cite{Kugo:1995km}.

Note that {\it e.g.} in maximally Abelian gauge the Kugo--Ojima scenario is not
applicable \cite{Hata:1992dn} and thus one does not expect the ghost propagator
to be IR enhanced. This has been recently confirmed in lattice
calculations \cite{Mendes:2006kc}.

\subsubsection{QCD Green functions: Violation of Positivity}

Given these considerations it is obvious that if states with transverse
gluons violate positivity these states do not belong to ${\rm Ker} Q_B$ and are
thus not physical states. One had to conclude that the transverse gluons belong
to a BRST quartet (together with a gluon-ghost, a gluon-antighost, and a
2-gluon state). Therefore it is sufficient to show positivity violation in the
gluon propagator to prove this kind of gluon confinement. 

At this stage it is interesting to point out that more than 25 years ago the
contradiction between antiscreening of gluons (and thus asymptotic freedom) on
the one hand and positivity of the gluon propagator on the other hand has been 
noted \cite{Oehme:1980ai}. In QCD in linear covariant gauges one obtains the 
following relation for the spectral sum rule of gluon correlation function
\begin{equation}
Z_3^{-1} = Z + \int_{m^2}^{\infty} d\kappa^2 \rho (\kappa^2)
  \quad {\rm with }\quad
  Z_3 = \left(
        \frac{g^2}{g_0^2}
        \right)^{\gamma}
\label{pos}
\end{equation}
where $Z$ is the one-particle contribution,
$Z_3$ is the gluon renormalization constant and $\gamma$ its corresponding
anomalous dimension. Due to antiscreening $Z_3^{-1} \rightarrow 0$, {\it i.e.}
$Z_3^{-1} \le Z$, and therefore one has to conclude that  $\rho (\kappa^2) \le
0$ for some values of $\kappa$. This means nothing else than positivity
violation in the gluon propagator. However, what is needed to compellingly
demonstrate such a positivity violation is to verify these arguments 
beyond perturbation theory.

There is another important issue which is also related to the question of
positivity: 
How is  the cluster decomposition theorem circumvented in Yang-Mills theory? 
Including the indefinite metric spaces of covariant gauge
theories this theorem can roughly be summarized as follows \cite{Haag:1996}:
For the vacuum expectation values of the local product of two field operators $A$
and $B$, being at a large spacelike separaration $R$ of each other, one
obtains the following bounds depending on the existence of a finite gap $M$ in
the spectrum of the mass operator \cite{Nakanishi:qm}
\begin{eqnarray}
&&        \Big|  \langle  \Omega | A(x) B(0) |\Omega \rangle  -
 \langle  \Omega | A(x) |\Omega \rangle  \,  \langle  \Omega
             |  B(0) |\Omega \rangle  \Big|  
    \le  \; 
\nonumber \\ 
&&\le  \;     \bigg\{ \begin{array}{ll}
   \mbox{\small Const.} \, \times \, R^{-3/2 + 2N} \, e^{-MR} \!\!, \quad
                      & \mbox{mass gap } M \; , \\
   \mbox{\small Const.} \, \times \, R^{-2 + 2N} \,, \;\;
                      & \mbox{no mass gap} \; ,  \end{array}  
\end{eqnarray}
for $R^2 = - x^2 \to \infty $. Herein, positivity entails that $N = 0$, but a
positive integer $N$ is possible for the indefinite state space 
of gauge theories. Therefore, in order to avoid the decomposition property for
products of unobservable operators $A$ and $B$ which together with the
Kugo-Ojima criterion is equivalent to avoiding the decomposition property for
colored clusters, there should be no mass gap in the indefinite space
$\mathcal{V}$. Of course, this implies nothing on the physical spectrum of the
mass operator in $\mathcal{H}$ which certainly should have a mass gap. In fact,
if the cluster decomposition property holds for a product $A(x) B(0)$ forming an
observable, it can be shown that both $A$ and $B$ are observables themselves.
This then eliminates the possibility of scattering a physical state into color
singlet states consisting of widely separated colored clusters (the
``behind-the-moon'' problem) \cite{Nakanishi:qm}. It has to be noted that the
Kugo-Ojima criterion implies the absence of a massless particle pole in  the
color charge operator, and therefore in $\partial^\nu F^a_{\mu\nu} $. This
shows that the unphysical massless ``excitations'' which are necessary to avoid
the cluster decomposition property are not the transverse gluons. 

In the following compelling evidence for the above described aspects of QCD 
in the covariant gauge, and thus for this kind of gluon confinement, will be
presented. It has to be emphasized, however, that this description is purely 
`kinematical', {\it i.e.\/} nothing is stated about the dynamics of
confinement.

\subsection{Infrared Exponents for Gluons and Ghosts}

\begin{figure}[htbp]
\begin{center}
\includegraphics[width=86mm]{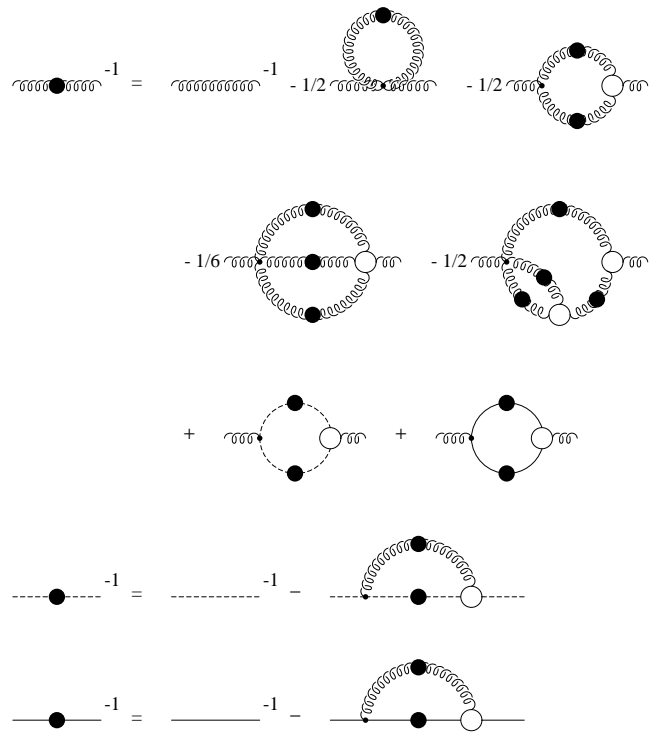}
\label{DSEs}
\caption{Diagrammatic representation of the propagator
           Dyson--Schwinger equations. The wiggly, dashed and solid
           lines represent the propagation of gluons, ghosts and quarks,
           respectively. A filled blob represents a full propagator and
           a circle indicates a one-particle irreducible vertex. }
\end{center}
\end{figure}

The tool which will be employed to study the above raised questions
is the set of Dyson--Schwinger equations (DSEs). The ones for the propagators of
Landau gauge QCD are displayed in fig.~3. One sees that the propagators couple
to higher $n$-point functions which are in general unknown. Thus the questions
arises: What can be infered from these equations? Is it possible to derive exact
results? Surprisingly the answer is: Yes, if one is willing to accept some
reasonable assumptions about the mathematical properties of Green functions.

\subsubsection{An exact inequality}

\begin{figure}[htbp]
\begin{center}
\includegraphics[width=36mm]{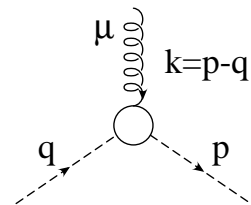}
\label{GGV}
\caption{Diagrammatic representation of the ghost-gluon vertex. }
\end{center}
\end{figure}

It will be shown that the general properties of the ghost
Dyson-Schwinger equation and one additional assumption, namely that QCD Green's
functions can be expanded in asymp\-totics series in the infrared, allow to prove
the Kugo-Ojima confinement criterion and the Gribov-Zwanziger condition
\cite{Watson:2001yv,Lerche:2002ep,Zwanziger:2001kw}.

The starting point is the non-renormalization of the ghost-gluon vertex in
Landau gauge to all orders in perturbation theory \cite{Taylor:1971ff}. It does
not acquire an independent ultraviolet renormalization, and even more, it stays
bare for vanishing out-going ghost momentum $p_\mu \to 0$, {\it c.f.\/} fig.~4.
These properties have been verified non-perturbatively
\cite{Cucchieri:2004sq,Schleifenbaum:2004id,Sternbeck:2005qj}. 
Due to this the ghost-gluon 
vertex cannot be singular for vanishing momenta which has important
consequences.

To fix the notation we note that in Landau gauge the gluon and ghost propagators 
are parametrized by
the two invariant functions $Z(k^2)$ and $G(k^2)$, respectively.
In Euclidean momentum space one has
\begin{eqnarray}
        D_{\mu\nu}(k) = \frac{Z(k^2)}{k^2} \, \left( \delta_{\mu\nu} -
        \frac{k_\mu k_\nu}{k^2} \right)  \; ,\quad
        D_G(k)  &=& - \frac{G(k^2)}{k^2}          \;.
	\nonumber \\
\end{eqnarray}
After renormalization these propagators depend also on the renormalization scale
$\mu$.
Furthermore assuming that the QCD Green functions can be expanded
in asymptotic series , {\it e.g.\ },
\begin{eqnarray}
G(p^2;\mu^2) = \sum _{n} d_n \left( \frac {p^2}{\mu^2} \right)^{\delta_n} ,
\end{eqnarray}
the integral in the ghost Dyson--Schwinger equation can be split in up in three
pieces. The infrared integral is too complicated to be treated analytically.
The ultraviolet integral, on the other hand, does not contribute to the infrared
behaviour. As a matter of fact, it is the resulting equation for the ghost
wave function renormalization which allows one to extract definite information
\cite{Watson:2001yv} without using any truncation or ansatz.

The results are that the infrared behaviour of the gluon and ghost propagators
is given by power laws, and that the exponents are uniquely related such that
the gluon exponent is -2 times the ghost exponent. As we will see later on this
implies an infrared fixed point for the corresponding running coupling. The signs
of the exponents are such that the gluon propagator is infrared suppressed as
compared to the one for a free particle, the ghost propagator is infrared
enhanced: The infrared exponent of the ghost propagator is negative, $\kappa <
0$. This exact inequality implies that the Kugo-Ojima confinement criterion and
the Gribov-Zwanziger condition are both fulfilled.

It is worth emphasizing that no truncation or specialized ansatz has been used
to obtain this result. The one employed assumption, namely that one is allowed
to expand Green functions in asymptotic series, is not considered to be
problematic. Besides this only the general structure of ghost DSE and
multiplicative renormalizibility have been used. The fact that the ghost-gluon
vertex is only subject to a finite renormalization turns out to be the property
which makes the tower of DSEs tractable.

\subsubsection{Infrared Expansion Scheme}

Given that the Yang-Mills propagators obey infrared power laws, can one
determine the infrared exponents of higher $n$-point functions? To this end the
corresponding $n$-point DSEs have been studied in skeleton expansion, {\it
i.e.} a loop expansion using dressed propagators and vertices, and an
asymptotic expansion has ben applied to all primitively divergent Green
functions  \cite{Alkofer:2004it}. It turns out that in this expansion the Green
functions can only be infrared singular in the kinematical limit where 
 all external momenta  go to zero.
Thus to determine the degree of possible singularities it is sufficient to
investigate the DSEs in the presence of only one external $p^2 \ll
\Lambda^2_{\tt QCD}$.  As an example consider the DSE for the three-gluon
vertex. In fig.~5 we see the full equation, in fig.~6 an approximation in the
lowest order of a skeleton expansion. In the presence of one (small) external
scale the approximated DSE has a selfconsistent power law solution given by
$(p^2)^{-3\kappa}$. Thus the vertex is strongly singular in the infrared.  One
can show by induction that this solution is also present if terms to arbitrary
high order in the skeleton expansion are taken into account. Therefore the
skeleton expansion provides the correct infrared solution of the DSEs.

\begin{figure}[htbp]
\begin{center}
\includegraphics[width=80mm]{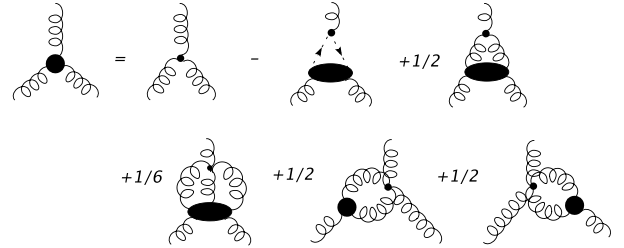}
\label{3gV}
\caption{Diagrammatic representation of the DSE for the 3-gluon vertex. }
\end{center}
\end{figure}

\begin{figure}[htbp]
\begin{center}
\includegraphics[width=80mm]{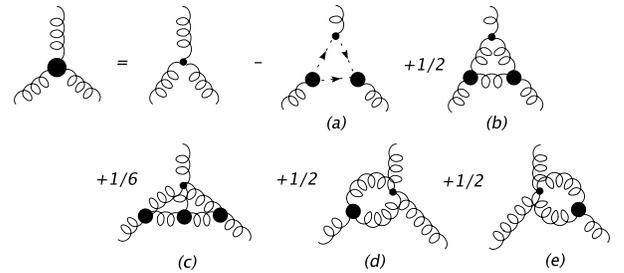}
\label{3gS}
\caption{Diagrammatic representation of the skeleton expansion 
for the 3-gluon vertex. }
\end{center}
\end{figure}

The following general infrared (IR) behaviour for one-particle irreducible
Green functions with $2n$ external ghost legs and $m$ external
gluon legs can be derived:
\begin{equation}
\Gamma^{n,m}(p^2) \sim (p^2)^{(n-m)\kappa}. \label{IRsolution}
\end{equation}
Very recently it has been shown \cite{Fischer:2006vf} by exploiting DSEs and
Exact Renormalization Group Equations that this IR solution is unique.
It especially includes that 
\begin{itemize}
\item 
{the ghost propagator is IR divergent.}
\item 
{the gluon propagator is IR suppressed.}
\item 
{the ghost-gluon vertex is IR finite.}
\item 
{the 3- and 4-gluon vertex are IR divergent if and only if all external momenta
vanish.}
\item 
{every coupling from an Yang-Mills vertex possesses an IR fixed point.}
\end{itemize}

\subsubsection{Numerical results from truncated equations}

The infrared solution described above verifies the infrared dominance of the
gauge fixing part of the covariantly gauge fixed QCD action conjectured in
ref.\  \cite{Zwanziger:2003cf}. The related infrared dominance of ghost loops 
and ultraviolet dominance of one-loop terms provides a reasoning for  earlier
used truncation schemes of DSEs being self-consistent at the level of two-point
functions \cite{vonSmekal:1997is}. These schemes have been refined and
generalized \cite{Fischer:2002hn} and allowed then to solve the coupled set of
DSEs for the ghost, gluon and quark propagators \cite{Fischer:2003rp}. Quarks
will be discussed in the next subsection, we discuss first the solution of the
truncated set of DSEs depicted in fig.~7.

\begin{figure}[htbp]
\begin{center}
\includegraphics[width=86mm]{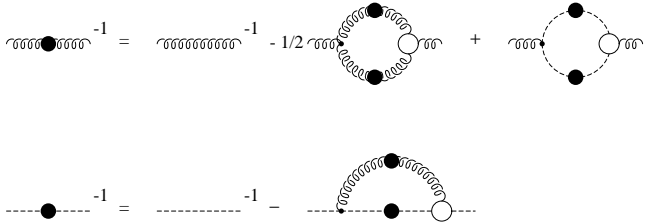}
\label{TruncDSE}
\caption{Diagrammatic representation of the numerically solved truncated DSEs. }
\end{center}
\end{figure}

As already noted the ghost DSE (taken in this truncation fully into account)
implies the IR behaviour ${G}(k^2) \rightarrow g(\kappa) (k^2)^{-\kappa}$
and ${Z}(k^2) \rightarrow f_1(\kappa) (k^2)^{2\kappa}$ whereas from the gluon
DSE one obtains  ${Z}(k^2) \rightarrow f_2(\kappa) (k^2)^{2\kappa}$.
As expected the ghost loop provides the infrared leading term.
Consistency then requires 
\begin{equation}
  f_1(\kappa) \stackrel{!}{=} f_2(\kappa)
  \quad \Rightarrow \quad
  \kappa = \frac{93 - \sqrt{1201}}{98} \simeq 0.595353.
\end{equation}
This result, first obtained from DSEs \cite{Lerche:2002ep,Zwanziger:2001kw},
has been verified using Exact Renormalization Group Equations
\cite{Pawlowski:2003hq,Fischer:2004uk}. 

As can be seen from refs.\ \cite{Fischer:2002hn,Fischer:2003rp} the numerical
results for the ghost and gluon propagators compare very well to corresponding
recent lattice data. However, the values of the infrared exponents extracted
from lattice calculations do neither agree with the analytical obtained DSE
results nor do they agree when compared against each other, see {\it e.g.\/}
refs.\ \cite{Oliveira:2006zg,MMP}. A comparison to lattice calculations
in three spacetime dimensions suggest that current lattice volumes are much too
small for reliable extraction of infrared exponents \cite{Maas:2006qw}.

At this point it is interesting to note that the DSEs can be solved on a compact
manifold with finite volume \cite{Fischer:2002hn,Fischer:2005ui}. A first study
of the volume dependence \cite{Fischer:2005ui} suggested that the continuum
limit would not be reached even at very large volumes. A recent investigation
\cite{Fischer:2006}, however, has shown that ultraviolet renormalization of the
DSEs is quite subtle when performed on a compact manifold. The correctly 
renormalized DSEs lead to numerical results which show the expected approach
to the continuum limit for increasing volumes \cite{Fischer:2006}.
This recent development will very likely facilitate the comparison of DSE and 
lattice results for QCD propagators in the future.

\subsection{Quark Propagator and\\ Dynamical Chiral Symmetry Breaking}

The Landau gauge quark propagator, $S(p)$,
in Euclidean momentum space can be generically written as
\begin{eqnarray}
  S(p) = \frac{1}{-i  p\!\!\!/\, A(p^2) + B(p^2)}
  =  \frac{Z_Q(p^2)}{-ip\hspace{-.5em}/\hspace{.15em}+M(p^2)}
  \, .
  \label{quark_prop}
\end{eqnarray}
A non-vanishing mass function, $M(p^2)$, for vanishing current quark mass
signals dynamical chiral symmetry breaking.

\begin{figure}[htbp]
\begin{center}
\includegraphics[width=80mm]{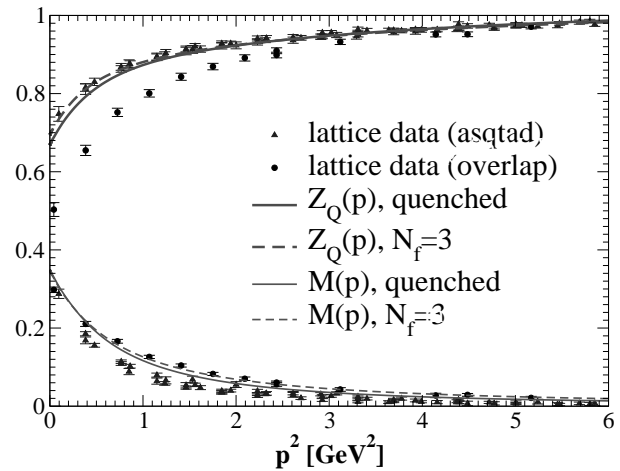}
\label{MLatt}
\caption{The dynamically generated quark mass function $M(p^2)$ and the quark
renormalization function $Z_Q(p^2)$ (both quenched and unquenched)
as obtained from DSEs in comparison to the chiral extra\-polation of 
quenched lattice
data \cite{Zhang:2003fa,Bowman:2002bm}, for details see text.}
\end{center}
\end{figure} 

Following the arguments for the generation of a confinement scale  given in the
introduction one easily also verifies that the  dynamical generation of quark
masses is also a genuinely non-perturbative  phenomenon. It furthermore 
requires a careful treatment of the quark-gluon interaction. In ref.
\cite{Fischer:2003rp} it is demonstrated that sizeable nontrivial
Dirac structures in the quark-gluon vertex are necessary to generate dynamical
quark masses of the order of 300-400 MeV. Our results for the quenched quark
mass  function $M(p^2)$ and the wave function $Z_Q(p^2)$ are compared to the
chiral extrapolations of the quenched lattice results of refs.
\cite{Zhang:2003fa,Bowman:2002bm} in fig.~8. 
Quite obviousy
the overall qualitative and quantitative  agreement bet\-ween both approaches 
is very good. The DSE results are within the bounds given by the two different
formulations of fermions on the lattice.

Including the backreaction of the quark-propagator on the ghost-gluon
system leads to a coupled set of three Dyson-Schwinger equations for the
propagators of  QCD. These equations have been solved in \cite{Fischer:2003rp}
and allowed a prediction of possible effects of unquenching QCD on the
propagators. Including $N_f=3$ chiral quarks
in the gluon DSE hardly changes the results for the quark propagator, see fig.\
8. 

Unquenched lattice results for the gluon propagator including the effects of
two light (up-) and one heavy (strange-) quark have been published recently
\cite{Bowman:2004jm}. In the gluon propagator the screening effect from
dynamical quarks  is clearly visible in the lattice results for momenta $p$
larger than $p=0.5$ GeV. This effect is also clearly
present in the DSE results, see {\it e.g.\/} ref.\ \cite{Fischer:2004wf}.

\subsection{Running Coupling}

The definition of a non-perturbative running coupling rests on a specifically
chosen Green function. As  the ghost-gluon vertex in Landau gauge acquires no
independent renormalisation, this relates the charge renormalisation constant
$Z_g$ to the ones for the gluon and ghost wave functions, 
\begin{equation}
  \widetilde{Z}_1 \, = \, Z_g Z_3^{1/2} \widetilde{Z}_3 \, = \, 1 \; ,
  \label{wtZ1}
\end{equation}
where the gluon leg provides a factor $\sqrt{Z_3}$,
the two ghost legs $\widetilde Z_3$. As we will demonstrate in the following
this allows for a definition of  a non-perturbative running coupling resting
solely on the properties of the gluon and ghost propagators.

>From the above relations between renormalization constants one concludes that 
$g^2 Z(k^2) G^2(k^2)$ is renormalization group (RG) invariant. In absence of
any dimensionful parameter this dimensionless product is therefore a function
of the running coupling $\bar g$,
\begin{equation}
  g^2 Z(k^2) G^2(k^2) = f( \bar{g}^2(t_k, g)) \; , \quad
  t_k = \frac{1}{2} \ln k^2/\mu^2 \; .
  \label{gbar}
\end{equation}
Here, the running coupling $\bar g(t,g)$ is the solution of
$d \bar g / dt = \beta(\bar g) $ with $\bar g(0,g) = g$ and the
Callan--Symanzik $\beta$-function $\beta (g) = - \beta_0 g^3 + {\mathcal
O}(g^5)$. The perturbative momentum subtraction scheme is asymptotically
defined by $f(x) \to x$ for $ x\to 0$. This is realized by independently
setting
\begin{equation}
  Z(\mu^2) = 1 \quad \mbox{and} \quad G(\mu^2) = 1
  \label{persub}
\end{equation}
for some asymptotically large subtraction point $k^2 = \mu^2$.
(Note, however, that requiring both conditions is only consistent at
asymptotically large $\mu^2$, see below.)
If the quantity
$g^2 Z(k^2) G^2(k^2)$ is to have a physical
meaning, {\it e.g.}, in terms of a potential between static colour sources, it
should be independent under changes $(g,\mu) \to (g',\mu')$ according to
the RG for arbitrary scales $\mu'$. Therefore,
\begin{equation}
  g^2 Z({\mu'}^2) G^2({\mu'}^2) \, \stackrel{!}{=} \,  {g'}^2 = \bar g^2(\ln
(\mu'/\mu) , g) \; , \label{gbar'}
\end{equation}
and, $f(x) \equiv x $, $ \forall x$. This can thus be adopted as a physically
sensible definition of a non-perturbative running coupling in the Landau
gauge.
In the present scheme 
it is not possible to realize $f(x) \equiv x$ by simply
extending the perturbative subtraction scheme (\ref{persub}) to arbitrary
values of the scale $\mu$, as this would imply a relation between the functions
$Z$ and $G$ which is inconsistent with the leading infrared behaviour
of the solutions.
For the two propagator
functions the condition (\ref{persub}) is in general too restrictive to be
used for arbitrary subtraction points. Extending the perturbative
subtraction scheme, one is only allowed to introduce functions of the coupling
such that
\begin{equation}
  Z(\mu^2) \, = \, f_A(g) \quad \hbox{and} \quad G(\mu^2) \, = \, f_G(g)
  \quad \hbox{with} \quad f_G^2 f_A \, = \, 1 \; ,
  \label{npsub}
\end{equation}
and the limits $ f_{A,\, G} \to 1 \, , \; g \to 0 $. Using this it is
straightforward to see that for $k^2 \not= \mu^2$ one has ($t_k = \frac 1 2
(\ln k^2/\mu^2)$),
\begin{eqnarray}
  Z(k^2) &=& \exp\bigg\{ -2 \int_g^{\bar g(t_k, g)} dl \,
  \frac{\gamma_A(l)}{\beta(l)} \bigg\} \, f_A(\bar g(t_k, g)) \; ,
  \label{RGsol} \\
   G(k^2) &=& \exp\bigg\{ -2 \int_g^{\bar g(t_k, g)} dl \,
  \frac{\gamma_G(l)}{\beta(l)} \bigg\} \, f_G(\bar g(t_k, g)) \; .
  \nonumber
\end{eqnarray}
Here $\gamma_A(g)$ and $\gamma_G(g)$ are the anomalous dimensions
of gluons and ghosts, respectively. 
Eq.\ (\ref{wtZ1}) corresponds to the following identity
for these scaling functions in Landau gauge:
\begin{equation}
  2 \gamma_G(g) \, +\, \gamma_A(g)  \, = \, -\frac{1}{g} \, \beta(g)  \; .
  \label{andim} 
\end{equation}
One thus verifies that the product $g^2 Z G^2$ indeed gives the running
coupling.
Therefore the  non-perturbative definition of this running coupling 
can be summarized as follows:
\begin{equation}
\alpha_S(k^2) =  \alpha_S(\mu^2) Z(k^2;\mu^2)G^2(k^2;\mu^2).
\end{equation}
The infrared behaviour of the ghost and gluon propagators implies
that the product $Z(k^2)G^2(k^2)$ goes to a constant in the infrared,
correspondingly we find an infrared fixed point of the running coupling:
\begin{equation}
\alpha_S(0)=\frac{2\pi}{3 N_c}
\frac{\Gamma(3-2\kappa)\Gamma(3+\kappa)\Gamma(1+\kappa)}{\Gamma^2(2-\kappa)
\Gamma(2\kappa)} \; 
.
\end{equation}
For the gauge group SU(3) the corresponding numerical value is
$\alpha_S(0) \approx 2.972$. Of course, this result depends on the employed
truncation scheme. In ref.\ \cite{Lerche:2002ep}, assuming the infrared dominance
of ghosts it has been shown that the tree-level vertex result
$\alpha_S(0) \approx 2.972$, among the general class
of dressed ghost-gluon vertices considered in the infrared,
provides the maximal value for $\alpha_S(0)$.
If the exponent $\kappa$ is chosen
in an interval between 0.5 and 0.7, as strongly suggested by lattice results,
one obtains $\alpha_S(0)> 2.5$ \cite{Lerche:2002ep}.

\subsection{Analytic properties of propagators}

\subsubsection{Positivity violation for the gluon propagator}

The  positivity violation of the (space-time) propagator of
transverse gluons as predicted by the Oehme--Zimmermann superconvergence
relation and 
corresponding to the Kugo--Ojima and Gribov--Zwanziger scenarios
has been a long-standing conjecture for which there is now
compelling evidence, see  {\it e.g.\/} ref.\ \cite{Alkofer:2003jj} and
references therein. The basic features underlying these gluon properties,
are the infrared suppression of correlations of transverse gluons and the
infrared enhancement of ghost correlations as discussed above.
A simple argument given by Zwanziger makes this obvious: An
IR vanishing gluon propagator implies  for the space-time gluon
propagator being the Fourier transform of the momentum space gluon propagator:
\begin{equation}
0=D_{gluon}(k^2=0) = \int d^4x \; \;\; \;  D_{gluon}(x) \ .
\end{equation}
\begin{figure}[htbp]
\begin{center}
\includegraphics[width=80mm]{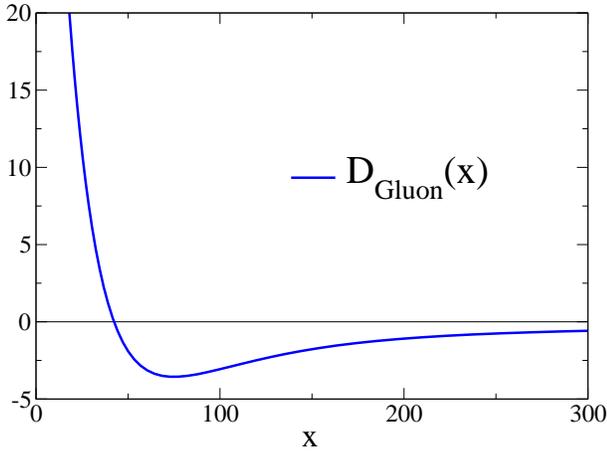}
\label{Gluon-pos}
\caption{The Fourier transform of the DSE result for the gluon propagator.}
\end{center}
\end{figure} 
This implies that $D_{gluon}(x)$ has to be negative for some values
of $x$.
And, as a matter of fact this behaviour is seen from fig.\ 9 in which 
the Fourier transform of the DSE result for the gluon propagator is displayed.

In order to investigate the analytic structure of the gluon propagator 
we first parameterize the running coupling such that the numerical results
for Euclidean scales are pointwise reproduced \cite{Fischer:2003rp}:
\begin{eqnarray}
\alpha_{\rm fit}(p^2) &=& \frac{\alpha_S(0)}{1+p^2/\Lambda^2_{\tt QCD}}\\
&+& \frac{4 \pi}{\beta_0} \frac{p^2}{\Lambda^2_{\tt QCD}+p^2}
\left(\frac{1}{\ln(p^2/\Lambda^2_{\tt QCD})}
- \frac{1}{p^2/\Lambda_{\tt QCD}^2 -1}\right) \nonumber 
\end{eqnarray}
with $\beta_0=(11N_c-2N_f)/3$. In this expression the Landau pole has been 
subtracted
({\it c.f.\/} ref.\ \cite{Shirkov:1997wi}), 
it is analytic in the complex $p^2$ plane except the real timelike axis
where the logarithm produces a cut for real $p^2<0$, and it obeys Cutkosky's rule.

The infrared exponent $\kappa$ is an irrational number, and thus  the
gluon propagator possesses a cut on the
negative real $p^2$ axis. It is possible to fit the solution for the gluon
propagator quite accurately without introducing further singularities
in the complex $p^2$ plane.
The fit to the gluon renormalization function \cite{Alkofer:2003jj}
\begin{equation}
Z_{\rm fit}(p^2) = w \left(\frac{p^2}{\Lambda^2_{\tt QCD}+p^2}\right)^{2 \kappa}
 \left( \alpha_{\rm fit}(p^2) \right)^{-\gamma}
 \label{fitII}
\end{equation}
is shown in Fig.~10. Hereby $w$ is a normalization parameter, and
$\gamma = (-13 N_c + 4 N_f)/(22 N_c - 4 N_f)$
is the one-loop value for
the anomalous dimension of the gluon propagator.
The discontinuity of (\ref{fitII}) along the cut
vanishes for $p^2\to 0^-$, diverges to $+\infty$ at $p^2=-\Lambda_{\tt QCD}^2$
and goes to zero for $p^2\to \infty$.

\begin{figure}[htbp]
\begin{center}
\includegraphics[width=86mm]{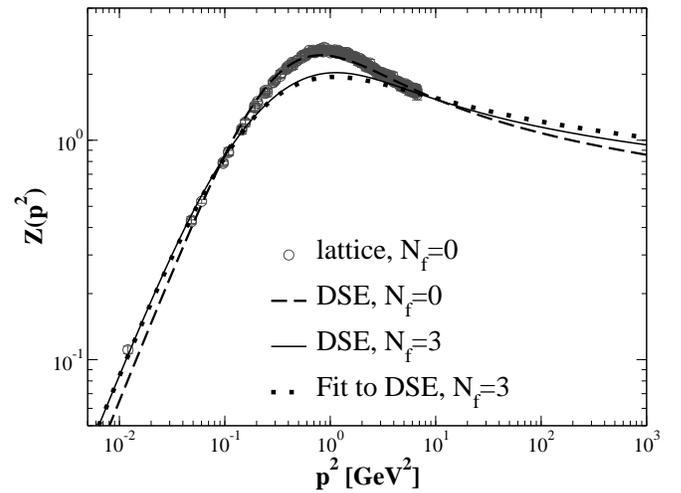}
\label{Gluon-pos2}
\caption{The gluon propagator 
compared to the fit, Eq.(\ref{fitII}), and lattice data \cite{Bonnet:2001uh}.}
\end{center}
\end{figure}

The function (\ref{fitII}) contains only four parameters:  the overall magnitude
which due to renormalization properties is arbitrary (it
is determined via the choice of the renormalization scale),   the
scale $\Lambda_{\tt QCD}$,   the infrared exponent $\kappa$ and  the
anomalous dimension of the gluon $\gamma$. The latter two are not
free parameters: $\kappa$ is determined from the infrared properties of the
DSEs and for $\gamma$ its one-loop value is used. Thus we have found a
parameterization of the gluon propagator which has effectively only one
parameter, the scale $\Lambda_{\tt QCD}$.
It is important to note that the gluon propagator possesses a form such 
that {\em Wick rotation is possible!}

Furthermore it is worth mentioning that the positivity violations for gluons is
also found at very high temperatures, even in the infinite temperature limit
\cite{Maas:2005hs,Cucchieri:2003di}. For the gluons being transverse to the
medium the Gribov-Zwanziger and/or Kugo-Ojima scenario  applies in the confined
and ``deconfined'' phases! This should not come as a real surprise: The
infinite temperature limit corresponds to three-dimensional Yang-Mills theory
plus an additional Higgs-type field, the left-over of the $A_4$ field. The
latter decouples in the IR, the three-dimensional Yang-Mills theory is as
expected confining and thus the corresponding gluon modes are positivity
violating.

\subsubsection{Analytic structure of the quark propagator}

>From the discussion above it is obvious that a dressed quark-gluon vertex is
mandatory. Especially those parts of the quark-gluon vertex reflecting 
dynamical chiral symmetry breaking are important. Their existence provides a
significant amount of self-consistent enhancement of dynamical mass generation.
In ref.\ \cite{Alkofer:2003jj} it has been found that fairly independently of
the form of the gluon propagator the resulting quark propagator respects
positivity if such scalar terms are of sufficient strength
in the quark-gluon vertex.   
The leading singularity of the quark propagator is then on the real axis, and 
the location of this singularity may play the role of a constituent quark mass.
The results of ref.\ \cite{Alkofer:2003jj}, however, strongly suggest that 
this singularity is not an isolated pole.

\subsection{Quark Confinement}

At this point of investigations gluon confinement is inherent but quark
confinement stays a mystery. Also, the precise structure of the quark
propagator depends crucially on the quark-gluon vertex, and therefore a
detailed study of this three-point function, and especially its IR behaviour,
is required to proceed.

To extend the above described IR analysis of Yang-Mills theory to full QCD
\cite{Alkofer:2006gz} one concentrates first on the quark sector of quenched
QCD and chooses the masses of the valence quarks to  be large, {\it i.e.\/}  $m
> \Lambda_{\tt QCD}$.   The remaining scales below $\Lambda_{\tt QCD}$ are
those of the external momenta of the Green functions. Without loss of
generality these can be  chosen to be equal, since infrared singularities in
the corresponding loop integrals  appear only when all external scales go to
zero.   One can then employ DSEs to determine the selfconsistent solutions  in
terms of powers of the small external momentum scale $p^2 \ll \Lambda_{\tt
QCD}$. The DSEs which have to be considered in addition to the DSEs of
Yang-Mills theory are the one for the quark propagator and the quark-gluon
vertex. The dressed quark-gluon vertex $\Gamma_\mu$ consists in general of
twelve linearly independent Dirac tensors. Especially those Dirac-scalar
structures are, in the chiral limit, generated non-perturbatively together with
the dynamical quark mass function in a self-consistent fashion: Dynamical
chiral symmetry breaking reflects itself, as anticipated by previous quark
propagator DSE results, thus not only in the propagator but also in a
three-point function. 

An IR analysis of the full set of DSEs reveals a non-trivial   solution for the
quark-gluon vertex: vector and {\em scalar} components of this  vertex are
infrared divergent with exponent $-\kappa - \frac 1 2$ \cite{Alkofer:2006gz}.
A numerical solution of a truncated set of DSEs confirms this infrared behavior.
Similar to the Yang-Mills sector the diagrams containing  ghost loops
dominate.  Thus all IR effects from the Yang-Mills sector are generated by
the IR asymp\-totic theory described above. More importantly, in the quark
sector the driving pieces of this solution are the scalar Dirac amplitude of the
quark-gluon vertex and the scalar part of the quark propagator. Both pieces are
only present when chiral symmetry is broken, either explicitely or dynamically.

\begin{figure}[htbp]
\begin{center}
\includegraphics[width=86mm]{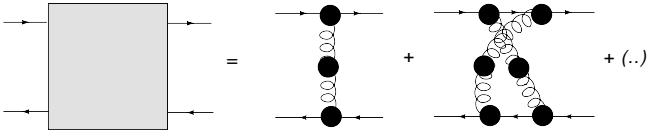}
\label{QC}
\caption{The four-quark 1PI Green's function and the first terms of its skeleton
expansion.}
\end{center}
\end{figure} 

The static quark potential is obtained from the four-quark 1PI Green's
function $H(p)$, whose skeleton  expansion is displayed in Fig.~11. From
its IR analysis one infers that  $H(p) \sim (p^2)^{-2}$for $p^2\to0$.
>From the well-known relation
\begin{equation}
V({\bf r}) = \int \frac{d^3p}{(2\pi)^3}  H(p^0=0,{\bf p})  e^{i {\bf p r}}
\ \ \sim \ \ |{\bf r} |
\end{equation}
between the static four-quark function $H(p^0=0,{\bf p})$ and the quark
potential $V({\bf r})$ one therefore obtains a linear rising potential.
Correspondingly, the running coupling from the quark-gluon vertex turns out to
be proportional to  $1/p^2$ in the infrared, {\it i.e.\/} contrary to the
couplings  from the Yang-Mills vertices this coupling is  singular in the
infrared.

The first term in the skeleton expansion, {\it i.e.\/} the effective, 
nonperturbative one-gluon exchange displayed in Fig.~11,  already generates
this result. Since most of the terms in the expansion are equally enhanced in
the IR, the string tension can only be calculated by summing over an
infinite number of diagrams. This property alleviates the usefulness
of the approach but it had to be expected in the first place. Since already an
effective, nonperturbative one-gluon exchange generates the confining potential
one is confronted with the problem of unwanted van-der-Waals forces. The
suppressed gluon propagator looks at first sight helpful because it implies
that there are  no long-range correlations between the gauge fields, and thus
no long-range correlations for chromoelctric and chromomagnetic fields at large
distances. However, the problem of avoiding long-range multipole fields has
only be shifted from the two-point correlation to the quark-gluon vertex.

\subsection{Summary of Part I}
The following points are the most important results described so far:
\begin{itemize}
\item[$\blacktriangleright$]
   Gluons are confined by ghosts,  positivity of transverse gluons is 
   violated.\\
   {Gluons are therefore removed from the $S$-matrix.}\\
   ({\it c.f.\/} the Kugo-Ojima confinement scenario, 
   the Oehme-Zimmer\-mann~superconvergence relation,
   the Gribov-Zwanziger horizon condition, etc.)
\item[$\blacktriangleright$]
   Chiral symmetry is dynamically broken\\ (in two- and {\bf three}-point 
   functions).
\item[$\blacktriangleright$]
 In the Yang-Mills sector the strong running coupling is IR finite.
\item[$\blacktriangleright$]
   The analytic structure of gluon and quark propagators is (very likely) 
   such that Wick rotation is possible.\\
   {$\bullet$ Effectively one parameter for the gluon propagator.}\\
   {$\bullet$ Evidence for constituent quark mass.}
\item[$\blacktriangleright$]
  There is compelling evidence that quark confinement in the Landau gauge is 
  due to the IR divergence of the quark-gluon vertex. 

  In the IR this vertex is dominated by its scalar components 
  thereby inducing a  
  relation between confinement and
  broken chiral symmetry.  
\end{itemize}

\section{Aspects of the Confinement mechanism in Coulomb-gauge QCD}

\subsection{Starting point}

As compared to the previous part of these lectures
the gauge will be changed from covariant to
Coulomb gauge. The aim is to relate the confinement of quarks to the
confinement of coloured composites \cite{Alkofer:2005ug}.

To this end we  start from the commonly accepted Wilson criterion
\cite{Wilson:1974sk} and an inequality between the gauge-invariant
quark-antiquark potential $V_W(R)$ and the color-Coulomb potential
$V_C(\vec{x})$  \cite{Zwanziger:2002sh}. The latter quantity is the
instantaneous part of the time-time component of the gluon propagator in
Coulomb gauge: $D_{00}(\vec{x},t)\propto V_C(\vec{x})\,\delta(t)+$
non-inst.~terms. In ref.~\cite{Zwanziger:2002sh} it was shown that if $V_W(R)$
is confining, {\it i.e.}~if $\lim_{R\rightarrow\infty}V_W(R)\rightarrow\infty$,
then also $|V_C(\vec{x})|$ is confining.
This was confirmed in SU(2) and SU(3) lattice calculations 
\cite{Greensite:2004ke,Nakamura:2005ux}
where it was found that $-V_C(\vec{x})$ rises linearly with $R=|\vec{x}|$.
However, the corresponding string tension, $\sigma_c$, was extracted to be
several times the asymp\-totic one, $\sqrt{\sigma_c}\approx$ 700MeV.

Furthermore, some of the basic features of Coulomb gauge QCD will be employed.
The presented investigation builds on properties of the gluon propagator 
in this gauge (see {\it e.g.\/}
refs.\ \cite{Szczepaniak:2003ve,Szczepaniak:2002wk,Szczepaniak:2001rg,
Zwanziger:2003de,Feuchter:2004mk}), investigations of the dynamical breaking of
chiral symmetry in corresponding Green function approaches (see {\it e.g.\/}
refs.\ \cite{Finger:1981gm,Govaerts:1983ft,Adler:1984ri,Alkofer:1988tc,
Llanes-Estrada:2004wr}) and related results of lattice calculations (see {\it
e.g.\/} refs.\
\cite{Greensite:2004ke,Cucchieri:2000gu,Greensite:2004bv,Greensite:2003xf,
Greensite:2003bk}). Especially the cancellations of IR
divergent expressions \cite{Alkofer:1988tc} will be exploited.

Here  an instantaneous approximation is used, and the effects of transverse
gluons are negelected. These approximations drastically simplify the
technicalities  involved in the calculations. However, some  of the physics
contained in the system is lost. The results are qualitatively, but not
quantitatively significant. We therefore refrain from using physical
dimensions, but instead present the quantities in all graphs in appropriate
units of the Coulomb string tension $\sigma_c$. The reason for the qualitative
reliability of the calculations is that the underlying symmetries of the theory
are incorporated in the model via Slavnov-Taylor or Ward-Takahashi identities. 

\subsection{Quark Dyson-Schwinger equation}

One immediate advantage of Coulomb gauge is the fact that it is possible
to perform all calculations in Minkowski space. In the employed approximations
the quark propagator fulfills the Dyson-Schwinger equation
\begin{equation}\label{qselfedetail}
i\;S^{-1}(p) = /\!\!\!\!\!p-m -
C_f\,6\pi\,\int\!\frac{d^4q}{(2\pi)^4}V_C(\vec{k})\,\gamma_0\,S(q)\,\gamma_0\;,
\end{equation}
where $\vec{k}=\vec{p}-\vec{q}$ and $C_f=(N_c^2-1)/(2 N_c)=4/3$ is the second 
Casimir invariant of the fundamental representation of the gauge group.
The $q_0$-integration in Eq.~(\ref{qselfedetail}) can be performed easily. 
One makes the Ansatz
\begin{equation}
S^{-1}(p):=-i(\gamma_0p_0-\vec{\gamma}\cdot\vec{p}\;C(p) - B(p))
\end{equation}
and obtains two coupled integral equations for the functions $B(p)$ and $C(p)$
\begin{eqnarray}\label{gapequdetaila}
B(p)&=&m+\int \frac{d^3q}{2\pi^2}\;V_C(k)
\frac{M(q)}{\tilde{\omega}(q)}\\\label{gapequdetailb}
C(p)&=&1+\frac{1}{p^2}\int \frac{d^3q}{2\pi^2}\;V_C(k)\;\vec{p}\cdot\vec{q}\;
\frac{1}{\tilde{\omega}(q)}\;,
\end{eqnarray}
where $m$ is the
current quark mass, $\tilde{\omega}(p):=\sqrt{M^2(p)+\vec p^2}$, and
$M(p):=B(p)/C(p)$ is the quark ``mass function''.
Its infrared behavior is a result of dynamical chiral symmetry breaking.  
In this Minkowski space formulation it can be directly used to define a 
constituent quark mass.

The Coulomb-gluon part $V_C(k)$ of the interaction
is chosen to be highly IR singular, 
\begin{equation}\label{pot}
V_C(k) = \frac{\sigma_c}{(\vec k^2)^2}\;,
\end{equation}
where $\sigma_c$ is the Coulomb string tension.
In practical calculations $V_C(k)$ is regulated by a parameter 
$\mu_\mathrm{IR}$ such that the
momentum dependence is modified to
\begin{equation}\label{mupot}
V_C(k) = \frac{\sigma_c}{(\vec k^2)^2}\rightarrow \frac{\sigma_c}
{(\vec k^2+\mu_\mathrm{IR}^2)^2}\;.
\end{equation}
In this fashion all quantities and observables become
$\mu_\mathrm{IR}$-dependent and one obtains the final result for some
$f(\mu_\mathrm{IR})$ by taking the limit $f=\lim_{\mu_\mathrm{IR}\rightarrow
0}f(\mu_\mathrm{IR})$.

\subsection{Bethe-Salpeter equation}

 A quark-antiquark bound state is described by the Bethe-Salpeter equation
(BSE), which in its homogeneous  form is written as (for clarity
Dirac, flavor, and color indices are neglected)
\begin{equation}\label{bse}
\Gamma(P,q) = \int d^4k\;K(q,k,P)\;S(k_+)\;\Gamma(P,k)\;S(k_-)\;,
\end{equation}
where $P$ and $q$ are the quark-antiquark pair's total and relative 
four-momenta, $\Gamma(P,q)$ is the bound state's
Bethe-Salpeter amplitude (BSA), $k_\pm = k\pm P/2$ are the individual 
quark- and antiquark-momenta, and
$K(q,k,P)$ is the quark-antiquark scattering kernel. Note that the quark
propagator appears as input into the BSE.
The axial-vector Ward-Takahashi identity is employed to  ensure that the 
kernels of the quark Dyson-Schwinger equation and
Bethe-Salpeter equations for pseudoscalar states are related in such a way 
that chiral symmetry and its
dynamical breaking are respected by the truncation.
Here, corresponding to the rainbow approximation in the quark 
Dyson-Schwinger equation we employ the ladder
approximation in the $qq$ scattering kernel in the BSE.
In particular this leads to the correct behavior of the pion mass as a
function of the current quark mass. Especially the pion mass vanishes 
in the chiral limit.

For pseudoscalar mesons and correspondingly scalar
diquarks the BSA can be characterized in terms of two scalar functions $h(p)$
and $g(p)$, which essentially are the coefficients of the pseudoscalar and 
axial-vector structures
in the BSA, for details, see ref.~\cite{Langfeld:1989en} and references therein.
{\it E.g.\/} the BSE (\ref{bse})  leads to the following equation
for the amplitude $h(p)$: 
\begin{eqnarray}\label{bsedetail}
h(p)= \frac 1{\omega(p)}
\int\! \frac{d^3q}{2\pi^2}\;V_C(k)\;
\left[h(q)+\frac{m_\pi^2}{4\, \omega(q)}g(q)\right]\;,
\end{eqnarray}
where $\omega(p) = C(p)\,\tilde{\omega}(p)$.
Note that the same type of IR divergent integral appears as in the 
case of the quark DSE. 

For vector mesons and axial-vector diquarks the BSA has four linearly
independent amplitudes. The construction of the four coupled integral equations
corresponding to the BSE is analogous to the pseudoscalar case.

\subsection{Cancelation of infrared divergencies}

The way the functions $B(p)$ and $C(p)$ diverge by containing a part
$I(p)\propto 1/\mu_{IR}$ makes the mass function well-defined:
\begin{equation}
M(p)=\frac{I(p)M(p)+B_{reg}(p)}{I(p)+C_{reg}(p)}=
         \frac{B_{reg}(p)}{C_{reg}(p)} .
\end{equation}	 
The fact that $M(p)$ stays finite has interesting consequences: The quark
propagator vanishes, {\it i.e.} {\em quarks are confined}, nevertheless the
colour singlet quark condensate stays well-defined \cite{Alkofer:1988tc}.
We will see that in other colour singlet quantities also the IR divergence
cancels.

For a bound state two different situations may arise:\\ 
(i) The energy of a
static $q\bar{q}$ state may be finite as the IR divergent  quark self-energies
cancel precisely against the IR divergent binding energy, or \\
(ii) this cancelation is incomplete, the state's energy is infinite, and it
is removed from the spectrum, {\it i.e.\/} confined.

\subsection{Numerical results}


\begin{figure}[htbp]
\begin{center}
\includegraphics[width=80mm]{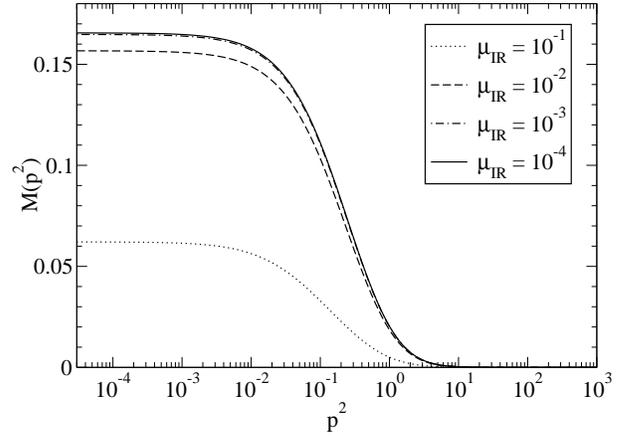}
\label{massfct2}
\caption{The quark mass function $M(q^2)$ for four values of the infrared 
regulator $\mu_\mathrm{IR}$
in the chiral limit $m=0$. All
quantities are given in appropriate units of $\sqrt{\sigma_c}$.}
\end{center}
\end{figure} 


\begin{figure}[htbp]
\begin{center}
\includegraphics[width=80mm]{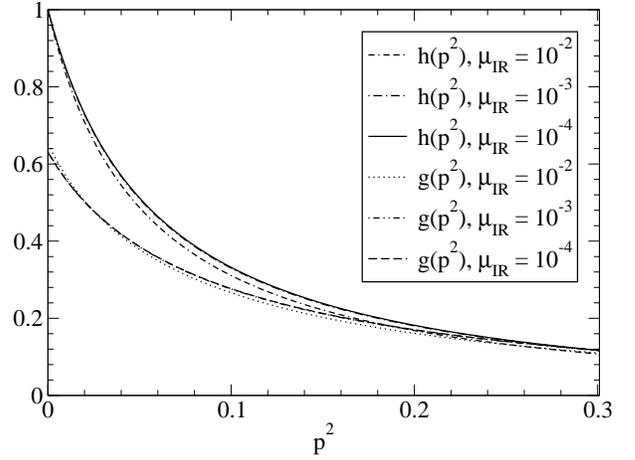}
\label{pimoments4}
\caption{Pion Bethe-Salpeter amplitude components $g$ and $h$ as functions of 
the infrared regulator $\mu_\mathrm{IR}$.
For convenience, the amplitudes are normalized such that $h(0)=1$.}
\end{center}
\end{figure} 

\begin{figure}[htbp]
\begin{center}
\includegraphics[width=80mm]{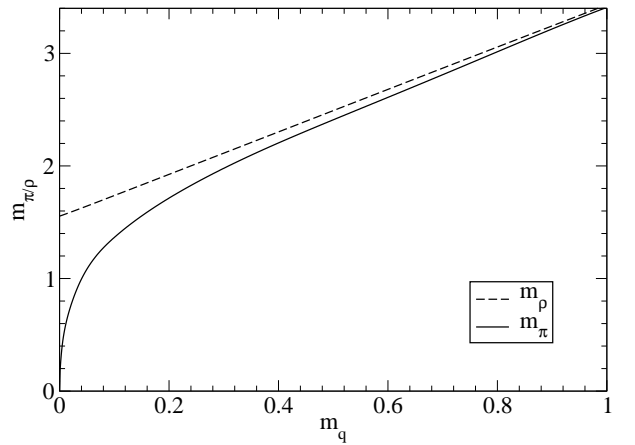}
\label{mesonmass1}
\caption{The pion and rho masses as functions 
of the current-quark mass in the limit $\mu_\mathrm{IR}\rightarrow 0$.}
\end{center}
\end{figure} 

In fig.~12 the quark mass function $M(q^2)$ is plotted for four different 
values of the infrared  regulator $\mu_\mathrm{IR}$ in the chiral limit $m=0$.
One nicely sees the convergence of the mass function.

The BSAs for mesons also behave as expected for  $\mu_\mathrm{IR}\rightarrow
0$. The results for the pion amplitudes $h(p)$ and $g(p)$ are
presented in fig.~13 (in an arbitrary normalization such that
$h(0)=1$).  As can be seen from fig.~14 the pion mass is vanishing in the 
chiral limit. Also the $\rho$ mass shows the expected dependence on the 
current quark mass.

\begin{figure}[htbp]
\begin{center}
\includegraphics[width=80mm]{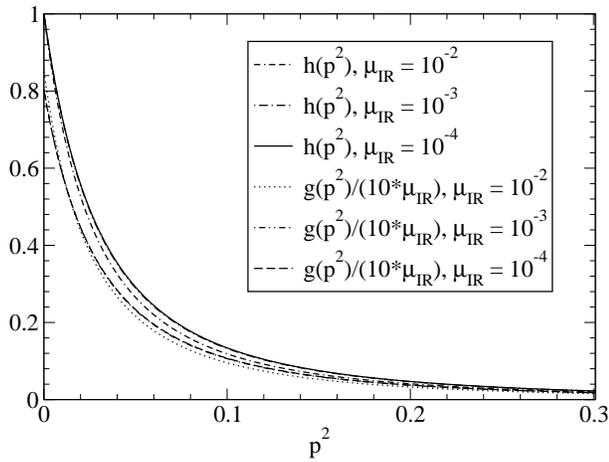}
\label{dimoments5}
\caption{Same as fig.~13 for the scalar diquark.}
\end{center}
\end{figure} 

\begin{figure}[htbp]
\begin{center}
\includegraphics[width=80mm]{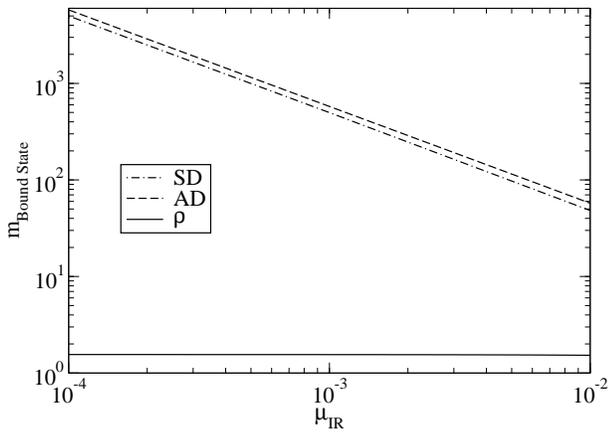}
\label{massmudep3}
\caption{The masses of the $\rho$ as well as the scalar (SD) and 
axial-vector (AD) diquarks as
functions of the infrared regulator $\mu_\mathrm{IR}$ in the chiral limit. 
The mass of the $\pi$ is identically zero for all
values of $\mu_\mathrm{IR}$ and therefore not shown in the graph.}
\end{center}
\end{figure} 

\begin{figure}[htbp]
\begin{center}
\includegraphics[width=80mm]{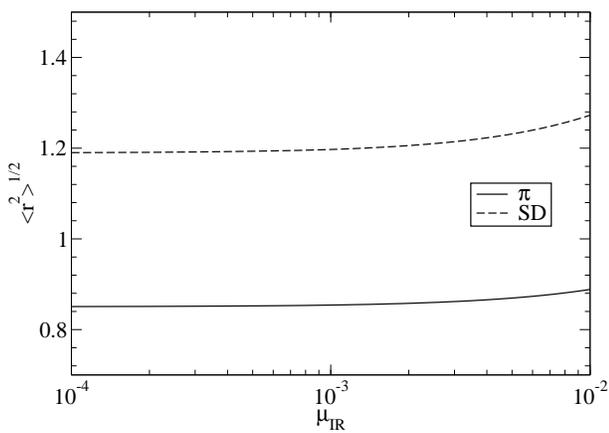}
\label{chargeradii6}
\caption{Charge radii for $\pi$ meson as well as scalar diquark (SD)
as functions of the infrared regulator $\mu_\mathrm{IR}$.}
\end{center}
\end{figure} 

The amplitudes for diquarks behave quite differently, the scalar diquark
amplitudes $h(p)$ and $g(p)$ are presented in fig.~15.
 IR cancelations appearing in the pion case
lead to a stable $h$ as well as ratio of $g/h$, which is not the case for 
the diquark: there
$g/h\sim \mu_\mathrm{IR}\rightarrow 0$ and $h\sim 1/\sqrt{\mu_\mathrm{IR}}$.
This has drastic consequences for the masses when plotted vs.\ the IR regulating
parameter $\mu_\mathrm{IR}$. In fig.~16 the chiral limit masses are displayed: 
one sees quite clearly that the $\rho$ mass is stable (the pion mass is anyhow
zero) but the diquark masses are diverging for vanishing $\mu_\mathrm{IR}$.
Therefore the diquarks are removed from the spectrum, {\em the diquarks are
confined}. 

It is then quite insructive to investigate the charge radii for both meson and
diquark states by calculating the electromagnetic form factors for small photon
virtualities. This gives finite results in the limit
$\mu_\mathrm{IR}\rightarrow 0$ for mesons and {\em diquarks}. Plots of the pion
and scalar diquark charge radii are shown in fig.~17. We conclude therefore that
the confined colour antitriplet quark-quark correlations possess a well-defined
size!

\subsection{Summary of Part II}

In Coulomb gauge QCD
\begin{itemize}
\item[$\blacktriangleright$]
the one-gluon-exchange of Coulomb-gluons ( {\it i.e.\/}
$D_{00}$) {\em over}confines!
\item[$\blacktriangleright$] the (four-dimensional) quark propagator vanishes,
 and quarks are confined!
\item[$\blacktriangleright$] dynamical chiral symmetry breaking occurs;\\
despite IR divergencies the colour singlet quark condensate is well-defined!
\item[$\blacktriangleright$] colour singlet meson properties are well-defined!
\item[$\blacktriangleright$] coloured diquarks are confined!
\item[$\blacktriangleright$]  they nevertheless possess a well-defined size!
\end{itemize}

\section{What is the nucleon?}

\subsection{Introduction}

The main objective of the studies to be  reported here is to develop a QCD
based understanding of the nucleon structure.  Recent experimental results
emphasize the complicated nature of baryons, and thus this aim is highly
ambitious. On the other hand, as seen from the previous parts of these lectures,
theoretical issues such as confinement, dynamical breaking of chiral symmetry
and the formation of relativistic bound states can be understood and related to
the properties of the non-perturbative propagators of QCD. This then can serve
as a basis to approach an {\it ab initio} calculation of nucleon
properties from continuum QCD.

A first step has been recently taken \cite{Eichmann:2006} by progressing
towards a bottom-up determination of the nucleons' quark core. Although some
technical limitations could have been yet only overcome by quite drastic
truncations it is worth to describe the corresponding first and quite
instructive results. Within the employed Poincar{\'e} covariant 
Dyson-Schwinger--Bethe-Salpeter--Faddeev approach the building blocks are as
realistic as currently available, however, yet not fully consistently
determined. As will become obvious the fact that meson, especially pion, cloud
effects are missing can nevertheless be clearly seen.

\subsubsection{A note on quark models}

It has to be emphasized here that in two important respects the presented
investigation differs from most quark model studies of the nucleon. First, the
resulting nucleon state is a four-momentum eigenstate. This is simply due to
the fact that Poincar{\'e} covariance has been respected in every step. Second,
there are no parameters adjusted to any observable. Starting with DSE or
lattice quark propagators this is simply not necessary. Especially, one inherits
the scale generated by dimensional transmutation when  non-perturbatively
renormalizing QCD.

Therefore it is not intended to add to the plethora of quark models like
soliton, bag or potential models. Over the last decades they have been useful
in gaining a phenomenological understanding on the experimental results of the
nucleon structure. 

The questions to be answered by the presented approach can be
considered being simple or being fundamental. Such questions are {\it e.g.}\\
~~~$\bullet$ How do baryons interact causally?\\
~~~$\bullet$ What r\^ole is played by the spin?\\
~~~$\bullet$ Is the nucleon spherically symmetric?\\
 Given the current situation in our understanding of the nucleon the ability to
answer  these questions can serve as a measure of the progress made 
in the investigation of hadron properties.


With these remarks in mind let us start by looking at properties of bound
states of the three Dirac fermions. 

\subsubsection{Relativistic angular momentum}

Even with relativistic valence quarks only, the nucleon has quite a rich
structure embodied in its wave function. This will be exemplified in the
nucleon's rest frame by a decomposition into partial waves w.r.t.\  the motion
of one of the valence quarks relative to the complementary  pair of quarks, see
{\it e.g.\/} ref.\ \cite{Alkofer:2005jh} and references therein. As  has been
demonstrated this analysis  also answers, without referring to a specific
dynamical model, the question whether the nucleon is spherically symmetric: It
is not -- due to the highly relativistic motion of quarks within the nucleon.

In non-relativistic physics angular momentum is defined w.r.t.\ a fixed origin.
Therefore the concept of angular momentum has to be generalized for
relativistic states. Mathematically one sees the effect of relativity from the
fact that the the Casimir operator of the non-relativistic rotation group,
$\vec J^2$, does not commute with boosts.

Describing the angular momentum with the help of a vector operator
orthogonal to the  particle momentum will cure the underlying problem. Thus we
will start our considerations from the Pauli-Lubanski axial-vector:
\begin{equation}
W_\mu = - \frac 1 2 \epsilon _{\mu\nu\rho\sigma } J^{\mu\nu} P^\sigma
\end{equation}
where  $J^{\mu\nu}$ is the Noether charge of rotations and boosts.
We note that
\begin{equation}
C_2=W_\mu W^\mu = m^2 j(j+1)
\end{equation}
is one of the two quadratic Casimir invariants of the Poincar\'e group, 
and that in the rest
frame it reduces to a quantity proportional to the usual spin:
\begin{equation}
W_\mu = (0,\vec W), \qquad W_i=- \frac 1 2 \epsilon _{ijk0} J^{jk} P^0
= - m \Sigma _i \; .
\end{equation}
Note that the decomposition into spin and angular momentum is {\em 
frame-dependent!}

Baryons, and as such also three-quark states, are eigenstates of parity, 
$P^2$ and
$W^2$. Coupling the three spin--$\frac{1}{2}$ quarks to a composite
spin--$\frac{1}{2}$  nucleon such that Poincar{\'e} covariance is maintained we
will see that \cite{Oettel:1998bk,Oettel:1999gc,Oettel:2000ig}:
\begin{itemize}
\item due to the compositeness, we need more components than four in total
or two for the positive energy states.
\item the lower components will not vanish in the rest frame
thus giving rise to the unavoidable presence of  at least a (relativistic)
$p$-wave contribution.
\item the difference of one angular momentum unit between upper and lower
components remains, however, there will be also a $d$-wave contribution.
\item the coupling to the electromagnetic field can be chosen such to maintain
causality, however, at the expense of a fairly complicated structure
of the nucleon--photon vertex containing one-- and  two-loop contributions.
\end{itemize}

\subsubsection{Quark-quark correlations}

In a baryon every quark pair is necessarily in a colour antitriplet state. In
this channel the interaction is attractive, certainly on a pertubative  level
and very likely also non-pertubatively. If we assume that in the corresponding
rest frame the angular momentum vanishes there remain only  two types of
states: Scalar (Spin 0) and axialvector (Spin 1) ``diquarks''. Note that these
are the analogous states to pseudoscalar and vector mesons because the intrinsic
parity of a fermion-fermion pair is opposite to the one of a
fermion-antifermion pair. For these diquark states the Pauli principle requires
flavour antisymmetry for the scalar and flavour symmetry for axialvector
quark-quark pair. 

\subsection{A Poincar{\'e}-covariant Faddeev Approach}

\subsubsection{Dyson-Schwinger equation}

The nucleon will appear as a pole in the quark six-point function. 
In this subsection a symbolic notation for all equations will be used,
otherwise it would be hard to recognize the relevant structures under all
the integrals and indices.
The corresponding fully renormalized Green function obeys a DSE 
\begin{equation}
G = G_0 + G_0 \,K \,G
  \quad \Longleftrightarrow \quad
  G^{-1} = G_0 ^{-1} - K
\end{equation}  
in terms of the tree-level six-point function $G_0$ and an interaction kernel
$K$, {\it c.f.\/} fig.~18. This equation, in Euclidean space, provides the
starting point for our investigations. In the vicinity of the pole the decomposition
$G\propto \bar \Psi \Psi /(P^2+m_N^2)$ yields
\begin{equation}
\Psi = G_0\,K\,\Psi
\end{equation}
which is diagrammatically displayed in fig.~19.

\begin{figure}[htbp]
\begin{center}
\includegraphics[width=80mm]{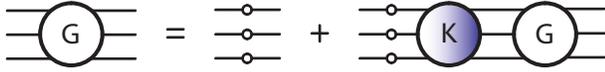}
\label{dse_nukleon}
\caption{A schematic representation of Dyson's equation for the quark 6-point
function.}
\end{center}
\end{figure} 

\begin{figure}[htbp]
\begin{center}
\includegraphics[width=50mm]{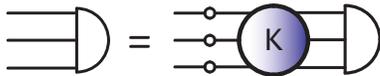}
\label{nukleon-bse}
\caption{The pole approximation to the DSE for the quark 6-point
function.}
\end{center}
\end{figure} 

QCD does of course also imply that irreducible three-particle interactions
exist between the three valence quarks of a nucleon. They are certainly
subleading in the ultraviolet which, however, is not the reason why we neglected
them here. In the previous lectures we have seen that there is a long-range
confining interaction between two quarks. We then simply assume that this force
is dominating the long-range correlation of also the three-quark state. When we
resort now in the next step to the Faddeev approximation, namely neglecting all
irreducible three-particle interactions (see fig.~20 for  a diagrammatic
representation) we  have to be aware that this may have a drastic influence on
our results. Currently this is a necessary step due to technical reasons. Only
if we are able to solve the  resulting Poincar{\'e}-covariant Faddeev equation,
see fig.~21, we will be able to overcome this, yet uncontroled, approximation.

Provided the Faddeev approximation is justified the kernel of the 6-point DSE
can be decomposed into three terms,
\begin{equation}
 K\,=\,K_1+K_2+K_3 \; .
 \label{k2pi}
\end{equation}
The {$K_i$}, $i= 1,2,3 $, describe the interactions of quark pairs $(jk)$, {\it
i.e.} with {quark ($i$)} as a spectator. $(ijk)$ is here a cyclic permutation
of (123). 
The two-quark propagators $g_i$ fulfill their own DSEs with kernels $K_i$,
\begin{equation}
 \label{g_i}
 g_i= G_0 +G_0 \; K_i \; g_i \; .
\end{equation}
The objects $g_i$  and $K_i$ are defined in three-quark space.
The former contain a factor $S_i$, the propagator of the spectator
quark, and the latter contain a factor of $S_i^{-1}$
(although the spectator quark is not involved in the interactions
described by $K_i$).
The two-quark scattering kernel $\widetilde{T}_i$ is defined
by amputating all incoming and outgoing quark legs
from the connected part of~$g_i$,
\begin{equation}
 g_i= G_0 + G_0 \; \widetilde{T}_i \; G_0 \; .
 \label{t1_def}
\end{equation}
Combining the two previous equations yields an integral equation
for $\widetilde{T}_i$,
\begin{equation}
 \widetilde{T}_i = K_i + K_i \; G_0 \; \widetilde{T}_i \; .
 \label{hatt_i}
\end{equation}
The so-called Faddeev components $\psi_i$ are introduced
by
\begin{equation}
 \Psi_i = G_0\; K_i \; \Psi \; ,
\end{equation}
which finally allows us to write down the Faddeev equation
\begin{equation}
 \Psi_i = S_j \,S_k \,\widetilde{T}_i \, (\Psi_j +\Psi_k) .
\label{Faddeev}
\end{equation}

\begin{figure}[htbp]
\begin{center}
\includegraphics[width=86mm]{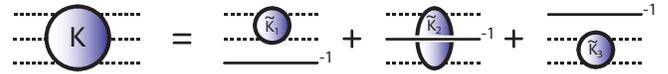}
\label{faddeevapproximation}
\caption{Diagrammatic representation of the Faddeev approximation.}
\end{center}
\end{figure} 

\begin{figure}[htbp]
\begin{center}
\includegraphics[width=86mm]{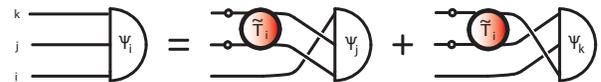}
\label{faddeevgln}
\caption{The Poincar{\'e}-covariant Faddeev equation of the nucleon.}
\end{center}
\end{figure} 

\subsubsection{Quark propagator}

In eq.\ (\ref{Faddeev}) the fully renormalized quark propagator appears
explicitely as well as implicitely (via the two-quark scattering kernel).  In
part I of these notes it is described how to obtain this quark propagator by
solving the Landau gauge DSEs for the quark, gluon and ghost propagators as
well as the quark-gluon vertex. In the calculations discussed below the
parameterization of ref.\ \cite{Alkofer:2003jj} with a cut on the real
time-like axis has been used. The quark DSE for complex
external momenta, as needed in  the Faddeev equation, is solved within an
on-going investigation \cite{Eichmann:2007}. However, these results have yet
not been included into a solution of the Faddeev equation.
There is another advantage  of using the parameterization of the DSE solution:
With some very mild adjustment of parameters it also describes to a high
accuracy the available lattice data for the  quark propagators. 

As has been discussed the dynamical chiral symmetry breaking, and thus the
dynamical generation of the infrared quark mass, is quantitatively well
described. The 'constituent quark' mass scale is described by $M(0) \approx 350
\dots 400$ MeV. The analytic structure is, on the other hand, quite sensitive
to the  details on the spacelike axis. In the chiral limit the location of the
singularity closest to the origin is $M(p_{sing}^2) \approx 500$~MeV when using
the DSE solution and 390 MeV when adjusting to lattice data. This then implies
finally a `range` of quark propagator parameterizations to be used.

As the parameterization of ref.\ \cite{Alkofer:2003jj} is based on DSE solutions
and/or lattice data with different values of the current mass the dependence of
the quark propagator functions on the current masses is kept quite precisely.
For some values of the current mass the constituent quark mass function is
plotted in fig.~22.

\begin{figure}[htbp]
\begin{center}
\includegraphics[width=86mm]{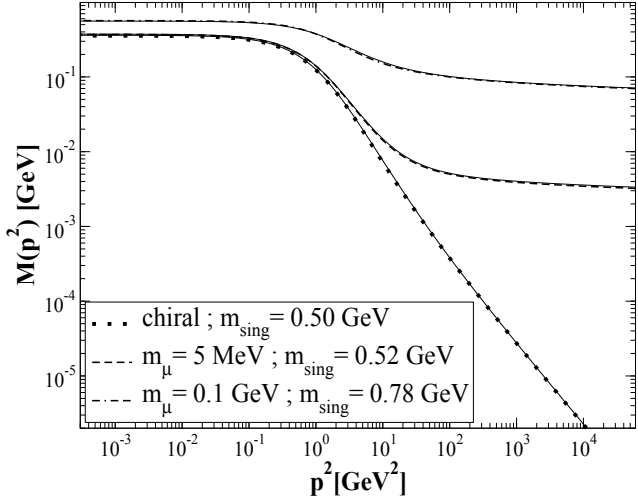}
\label{massfunction}
\caption{The mass functions of the employed quark propagators for different
values of the current mass calculated from the parameterization of ref.\
\cite{Alkofer:2003jj}.}
\end{center}
\end{figure}

\subsubsection{``Diquarks''}

In an on-going investigation \cite{Eichmann:2007}
the two-quark scattering kernel $\widetilde{T}_i$
is calculated for the complex momenta needed in the Faddeev equation. 
The results presented here are, however, calculated using a further
approximation, namely  by  representing the two-quark correlation function in
terms of a sum over separable correlations,
\begin{equation}
\widetilde{T}_i = \sum\limits_a
  \chi_i^a\,D_i^{\,a}\,\bar{\chi}_i^a \; ,
\end{equation}
which is pictorially shown in fig.~23.
\begin{figure}[htbp]
\begin{center}
\includegraphics[width=80mm]{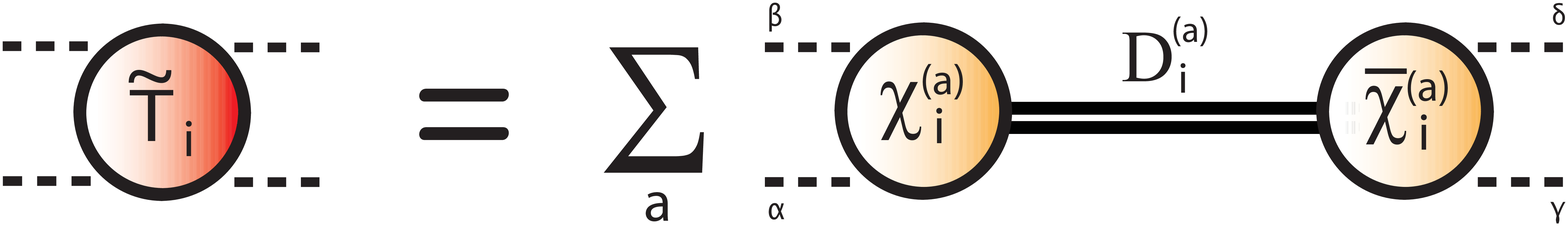}
\label{diquarks}
\caption{Diagrammatic representation of the separable approximation to the
two-quark scattering kernel $\widetilde{T}_i$.}
\end{center}
\end{figure} 
Hereby $D^a$ denotes the diquark propagator and $\chi^a,\bar{\chi}^a$ the
diquark amplitudes. The sum extends over scalar and axial-vector correlations.
For the amplitudes we take into account the corresponding leading Dirac
structures
\begin{eqnarray}
  \chi^5(p^2) &=& V_{\mbox{\tiny sc}}(p^2)\,\gamma^5 \,C ,\\
  \chi^\mu(p^2) &=& V_{\mbox{\tiny ax}}(p^2)\,\gamma^\mu \, C,
\end{eqnarray}
with $C$ being the charge conjugation Dirac matrix. The amplitudes $V(p^2)$,
reflecting the non-pointlike structure of quark-quark correlations, are
determined from an `on-shell` Bethe-Salpeter equation for the quark pair.
In this way the confined nature of ``diquarks'' is not taken into account,
and one obtains masses for the ``diquarks''. In the chiral limit, depending
on the quark propagator parameterization, one obtains
$m_{sc}\approx 0.6-0.8$ GeV,     $m_{ax}\approx 0.85-1$ GeV.
The quark pole mass and the diquark masses as a function of the current mass are
plotted in fig.~24.

\begin{figure}[htbp]
\begin{center}
\includegraphics[width=80mm]{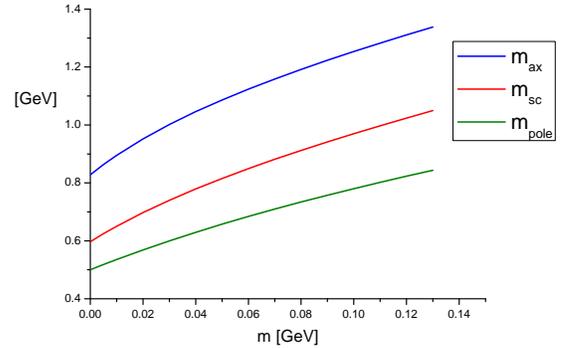}
\label{massensweep-dq}
\caption{Displayed are the quark pole mass and the diquark masses as a function
of the current mass for parameterization of the quark
propagator as resulting from coupled DSEs.}
\end{center}
\end{figure} 

\begin{figure}[htbp]
\begin{center}
\includegraphics[width=80mm]{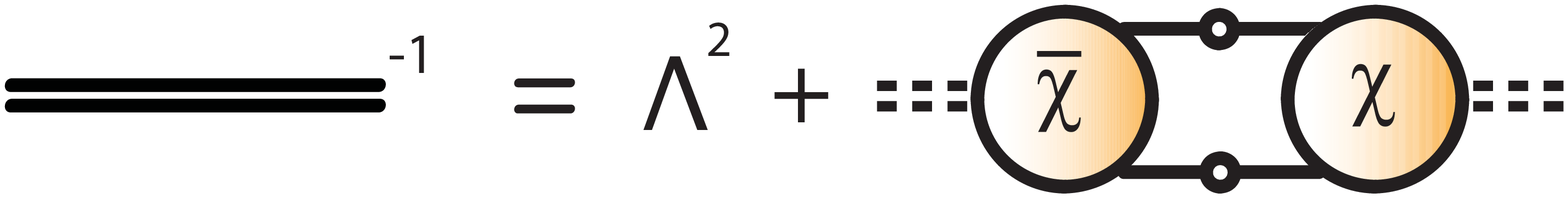}
\label{dqprop2.eps}
\caption{Reconstruction of the diquark propagator from a quark loop.}
\end{center}
\end{figure} 

Given the knowledge about the diquark BSAs and the diquark masses the diquark
propagator can be quite accuratelly represented by the sum of a constant and a
quark loop, see fig.~25. The constant is hereby chosen such that the correct 
``diquark msss'' is reproduced.

\subsubsection{Three-quark amplitudes}

Plugging the separable approximation for the  two-quark scattering kernel
$\widetilde{T}_i$ into the Faddeev equation leads to a coupled  set of 
quark-diquark Bethe-Salpeter equations which is best formulated for the
the vertex functions   $\Phi^{ab}$ related to the BSAs via
\begin{equation}
 \Psi^{ac}(p,P) = S(p_q)\; D^{ab}(p_d)\; \Phi^{bc}(p,P) \; .
 \label{wavedeldef}
\end{equation}
The quark-diquark Bethe-Salpeter equations then read:
\begin{equation}
  \phi^a(p,P)=\displaystyle\sum\limits_{b,c} \int\frac{d^4
  k}{(2\pi)^4}
  \underbrace{\chi^b\,S^T(q)\,\left.\bar{\chi}^a\right.^T}
  _{\displaystyle K_{BS}(p,k,P)} S(k_q)\,D^{b
  c}(k_d)\,\phi^c(k,P) \; .
\end{equation}

\begin{figure}[htbp]
\begin{center}
\includegraphics[width=86mm]{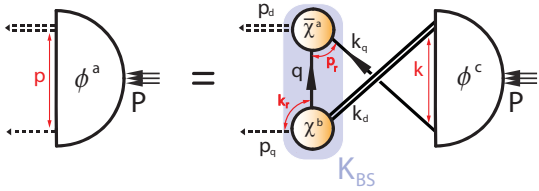}
\label{q-dq-bse}
\caption{Diagrammatic representation of the quark-diquark Bethe-Salpeter 
equations.}
\end{center}
\end{figure}

The interaction between the spectator quark and the correlated quark-quark pair
is quark exchange, which is also the least interaction required to reinstate
the Pauli principle. This interaction is attractive as antisymmetrisation in
colour requires the symmetrisation in all other quantum numbers.

The Bethe-Salpeter vertex function for the nucleon can be decomposed into
\begin{equation}
\Phi^N_{\alpha\beta\gamma} = \Phi^5_\alpha\,\, {
(\chi^5)_{\beta\gamma}}+
       \Phi^\mu_\alpha\,\, {(\chi^\mu)_{\beta\gamma}}
\end{equation}
\centerline{$
\begin{array}{lcl}
\Phi^5 &=& \sum_{i=1}^2 { S_i (p;P)}\,\,
    { \Gamma_i (\gamma^\mu,p,P)}\,\, u \\
 & & \\
\Phi^\mu &=& \sum_{i=1}^6 { A_i (p;P)}\,\,
    { \Gamma^\mu_i (\gamma^\mu,p,P)}\,\, u
\end{array}
$}

~\\ 
with constraints on the Dirac matrices $\Gamma_i$ such that
\begin{itemize}
 \item[] \hspace{-5mm} the nucleon has spin $\frac 1 2$, positive parity and positive energy.
 \item[] \hspace{-5mm} the two independent momenta
      \begin{itemize}
      \item[{ $P$}] =  baryon momentum, and 
      \item[{ $p$}] =  quark-diquark relative momentum,
      \end{itemize}
      \hspace{-5mm} are basis vectors.
\end{itemize}

\begin{table}
{ \small
  \begin{tabular}{|l|c|c|}
N wave fcts in the rest frame & eigenvalue & eigenvalue \\
  & $l(l+1)$ of ${\bf L}^2$  & $s(s+1)$ of ${\bf S}^2$ \\
   \hline \hline
  & & \\
\footnotesize{${\cal S}_1 u (\gamma_5 C)= {\chi \choose 0 }(\gamma_5 C)$}
 & 0 {s} & $\frac{3}{4}$ \\
 ~~~~~~~~~~~~~~~~~~~~~~~~~~~~~~{scalar}&& \\
\footnotesize{${\cal S}_2 u (\gamma_5 C)={0\choose
\frac{1}{p}(\vec{\sigma}\vec{p})\chi }(\gamma_5 C)$}
 & 2 {p}& $\frac{3}{4}$ \\
 \hline
\footnotesize{${\cal A}^\mu_{1} u (\gamma^\mu C)=\hat
P^0{\frac{1}{p}(\vec{\sigma}\vec{p})\chi\choose 0}(\gamma^4 C)$}
 & 2 {p} &$\frac{3}{4}$\\
 &&\\
\footnotesize{${\cal A}^\mu_{2} u (\gamma^\mu C)=\hat P^0{0\choose
\chi}(\gamma^4 C)$}
 & 0 {s} & $\frac{3}{4}$ \\
 & & \\
\footnotesize{${\cal B}^\mu_{1} u (\gamma^\mu C)={i\sigma^i\chi \choose 0}
(\gamma^i C)$}
 & 0 {s} & $\frac{3}{4}$ \\
 ~~~~~~~~~~~~~~~~~~~~~~~~~{axialvector}&&\\
\footnotesize{${\cal B}^\mu_{2} u (\gamma^\mu
C)={0\choose\frac{i}{p}\sigma^i(\vec{\sigma}\vec{p})\chi}(\gamma^i C)$}
 & 2 {p} & $\frac{3}{4}$ \\
 & & \\
\footnotesize{${\cal C}^\mu_{1} u (\gamma^\mu C)={i\left(\hat
p^i(\vec{\sigma}\hat{\vec{p}})-\frac{1}{3}\sigma^i\right)\chi\choose
         0}(\gamma^i C)$}
 & 6 {d}& $\frac{15}{4}$ \\
 &&\\
\footnotesize{${\cal C}^\mu_{2} u (\gamma^\mu C)={0\choose \frac{i}{p}\left(p^i-\frac{1}{3}\sigma^i(\vec{\sigma}\vec{p})\right)\chi}
               (\gamma^i C)$}
          & 2 {p} & $\frac{15}{4}$ \\
\hline
\end{tabular}}
\caption{The partial wave decomposition of the nucleon in its rest frame.}
\end{table}

\begin{figure}[htbp]
\begin{center}
\includegraphics[width=86mm]{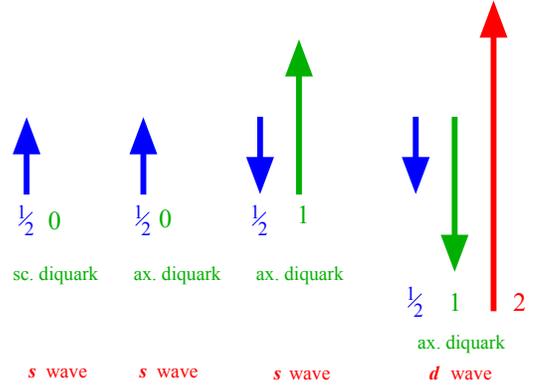}
\label{decomposition}
\caption{Angular momentum coupling of the nucleon's three valence quarks.}
\end{center}
\end{figure}  

The partial wave decomposition of the nucleon in the rest frame is given in
table~1. Besides three $s$--waves (two of them also present in the
non-relativistic limit) some of the lower components provide $p$--waves 
(four in total), and there is also a $d$--wave component. The existence of the
latter means that the nucleon is, as a quantum state of course only 
in the internal frame,
not spherically symmetric.  The angular momentum coupling is depicted in
fig.~27.

\subsubsection{Nucleon mass}

\begin{figure}[htbp]
\begin{center}
\includegraphics[width=80mm]{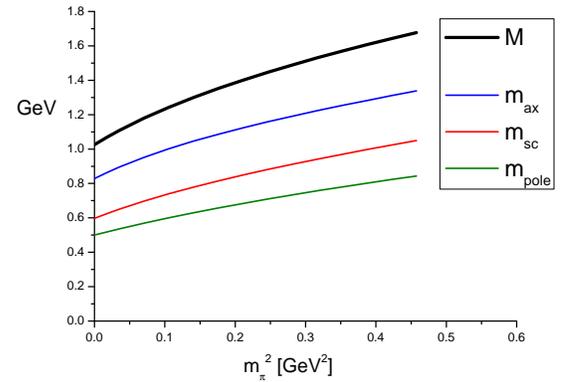}
\label{massensweep}
\caption{The mass attributed to the nucleon's quark core, the diquark
masses and the quark `pole' mass as function of the pion mass squared.}
\end{center}
\end{figure}

Solving the quark-diquark BSEs (for details of the numerical method see ref.\
\cite{Oettel:2001kd}) one obtains a mass which can be attributed to the
nucleon's quark core. This mass is above the physical value $M=0.94$ GeV.
Depending on the parameterization of the quark propagator and the calculated
diquark masses the chiral limit value ranges between $M \sim 1.07$ GeV $\dots
1.25$ GeV. In fig.~28 this mass is displayed as a function of the pion mass
squared together with diquark masses and the quark `pole' mass. From fig.~29 it
is evident that this overestimation of the nucleon mass is a result for all
values of the current quark mass (or pion mass, respectively). The comparison
to the results from Chiral Perturbation Theory and to lattice data
\cite{Gockeler:2003ay} clearly shows that we missed some attractive
contribution in the nucleon. A possible explanation which attributes this to
missing pion effects can be found in refs.\ \cite{Hecht:2002ej,Oettel:2002cw}.

\begin{figure}[htbp]
\begin{center}
\includegraphics[width=70mm]{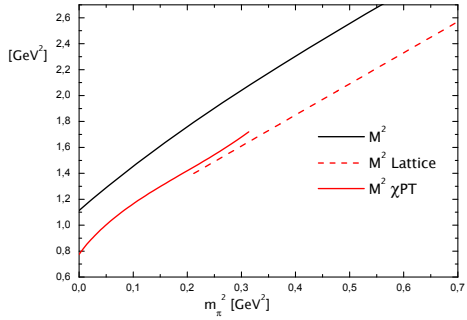}
\label{nukleonmasse}
\caption{The mass attributed to the nucleon's quark core
compared to results from Chiral Perturbation Theory and lattice data
\cite{Gockeler:2003ay} as function of the pion mass squared.}
\end{center}
\end{figure}

\subsection{Electromagnetic Current}

\begin{figure*}[htbp]
\begin{center}
\includegraphics[width=110mm]{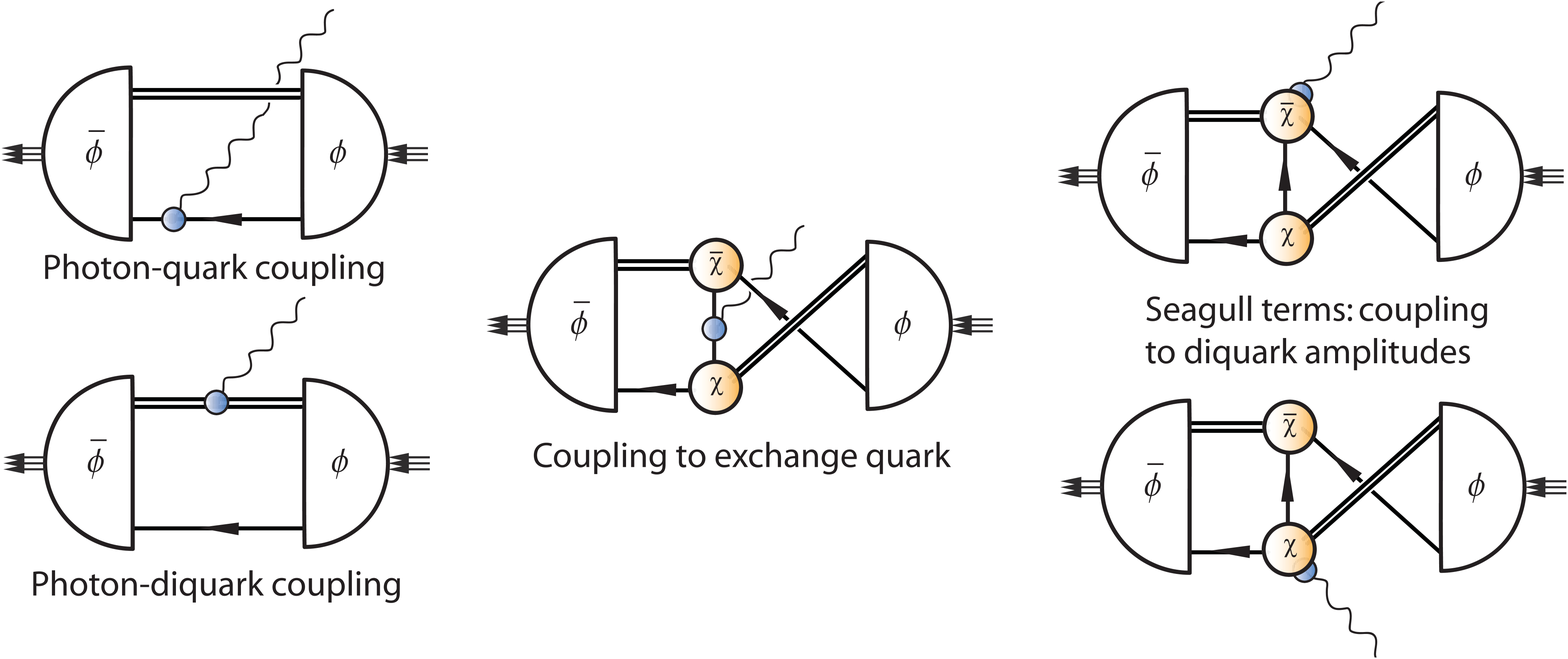}
\label{strom}
\caption{Feyman diagrams which are taken into account in the construction
of the nucleon's electromagnetic current operator.}
\end{center}
\end{figure*}

\subsubsection{Current Contributions}

To describe the coupling of the photon to the nucleon's quark core one has to
construct the electromagnetic current operator. The corresponding five-point
function has to fulfill a Ward-Takahashi identity which in turn guarantees that
current conservation holds 
\cite{Kvinikhidze:1999xn,Oettel:1999gc}. This requires 
to include at least the following diagrams (see fig.~30):
  \begin{itemize}
  \item photon-quark coupling,
  \item photon-diquark coupling,
  \item coupling to exchange quark,
  \item seagull terms: coupling to diquark amplitudes.
  \end{itemize}
Unknown constants like the anomalous magnetic moment
of the axialvector diquark and the strength of the
scalar-axialvector transitions will be calculated by taking into account
the diquark substructure  such that
gauge invariance for the nucleon current is respected \cite{Oettel:2000jj}.
Due to the non-trivial momentum dependence of the quark propagator the 
quark-photon vertex possesses a quite non-trivial  structure 
\cite{Ball:1980ay,Curtis:1990zs,Maris:2000bh}.
This then also implies a rich structure of the diquark-photon vertex,
for the construction of the latter and its use within a covariant 
Faddeev approach see 
{\it e.g.} refs.\ \cite{Weiss:1993kv,Ahlig:2000qu,Oettel:2002wf,Alkofer:2004yf}.

\subsubsection{Results on Electromagnetic Form Factors}

Equipped with the explicit form of the current operator the electromagnetic
form factors can be calculated. For the corresponding results the reader is
refered to \cite{Eichmann:2006}, here the focus will be on derived quantities
like the electric and magnetic radii as well as the magnetic moments.

>From table~II one sees that the nucleons' radii are underestimated,
hereby the magnetic radii more than the electric ones. One can test the
hypothesis whether this shortcoming is due to the missing of the pion cloud.
First of all, in the isoscalar combination of the radii pion effects are of the
second order and therefore small. A quantitative description of the nucleon's
quark core should give values close to the experimental ones, and this is what
we find. Second, at large pion mass the pion cloud effects should become less
important, and our results should agree with lattice data. As can be seen from
fig.~31 the agreement for the isovector part of the $F_2$ and $F_1$ radius is
certainly not perfect but our results are not too far off the lattice data.

\begin{figure*}[htbp]
\begin{center}
\includegraphics[width=140mm]{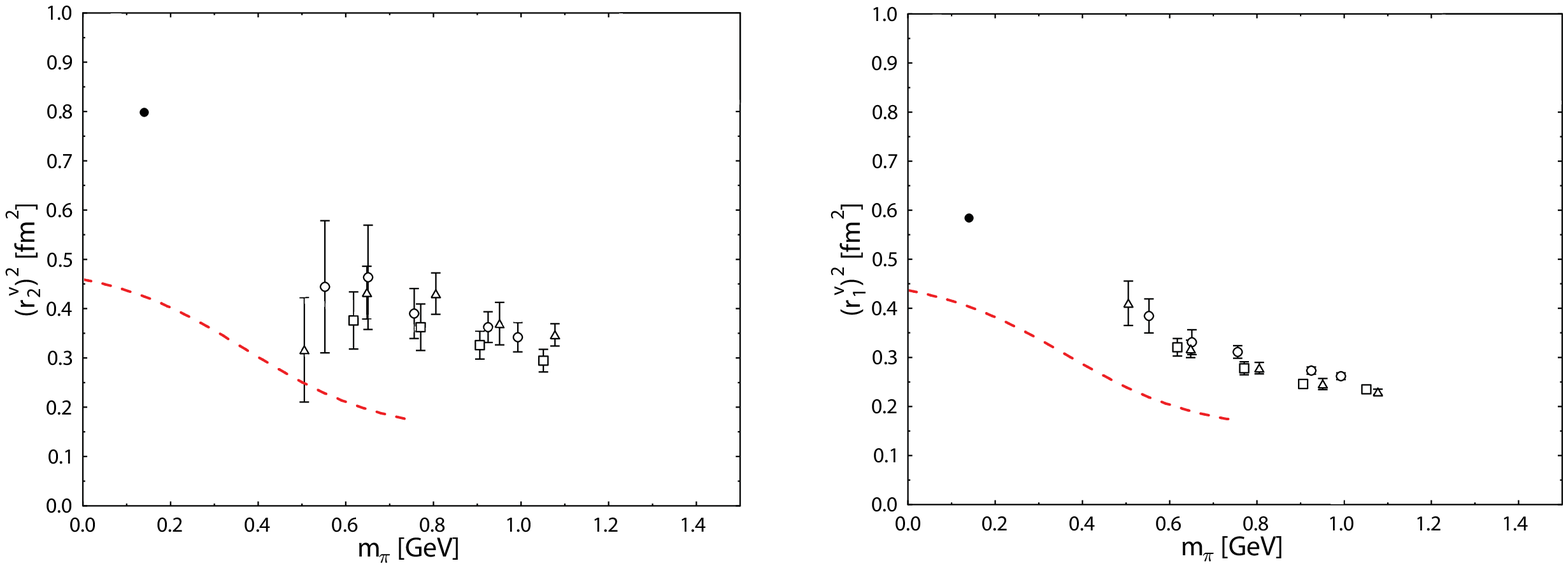}
\label{radii}
\caption{The results for the $F_2$ isovector radius (left panel) 
and the $F_1$ isovector radius (right panel) as a function of the pion mass
in comparison to the experimental value and corresponding lattice data
\cite{Gockeler:2003ay}.}
\end{center}
\end{figure*}

 \begin{table}
  \centering
  \begin{tabular}{l||c|c}
          & ~~~~~~~~~calc. ~~~~~~~~~& ~~~~~~~~~ {exp. }~~~~~~~~~ \\  
	  \hline\hline &&\\
  $r_E^p$ &  $0.73$ & {$0.836$} \\
  $r_E^n$ &  $0.29$ & {$0.336$} \\
  $r_M^p$ &  $0.64$ & {$0.843$} \\
  $r_M^n$ &  $0.60$ & {$0.840$} \\
  \end{tabular}
  \caption{The nucleons' electromagnetic radii (in fm) as calculated 
  in the covariant Faddeev approach in comparison to experimental values.}
  \end{table}

  \begin{table}
  \centering
  \begin{tabular}{l||c|c|c}
          & ~~~~~~~~~calc. ~~~~~~~~~ &  ~~~~~~~~~ {exp.} ~~~~~~~~~ & 
	  without pions \cite{Hemmert:2002uh}\\  
	  \hline\hline
	  &&\\
  $\mu_p$ &  $ ~~3.62$ & {$ ~~2.793$} & $  ~~3.5 \pm 0.2 $ \\
  $\mu_n$ &  $-2.34$ & {$-1.913$} & $ -2.6 \pm 0.2 $ \\
  $\kappa_s$ &  $ ~~0.28$ & {$ -0.120$} & $  -0.11$ \\
  $\kappa_v$ &  $ ~~4.96$ & {$ ~~3.706$} & $  ~~5.1 \pm 0.4   $ \\
  \end{tabular}
  \caption{The nucleons' magnetic moments (in n.m.) as calculated 
  in the covariant Faddeev approach in comparison to experimental values.}
  \end{table}

\begin{figure}[htbp]
\begin{center}
\includegraphics[width=8.4cm]{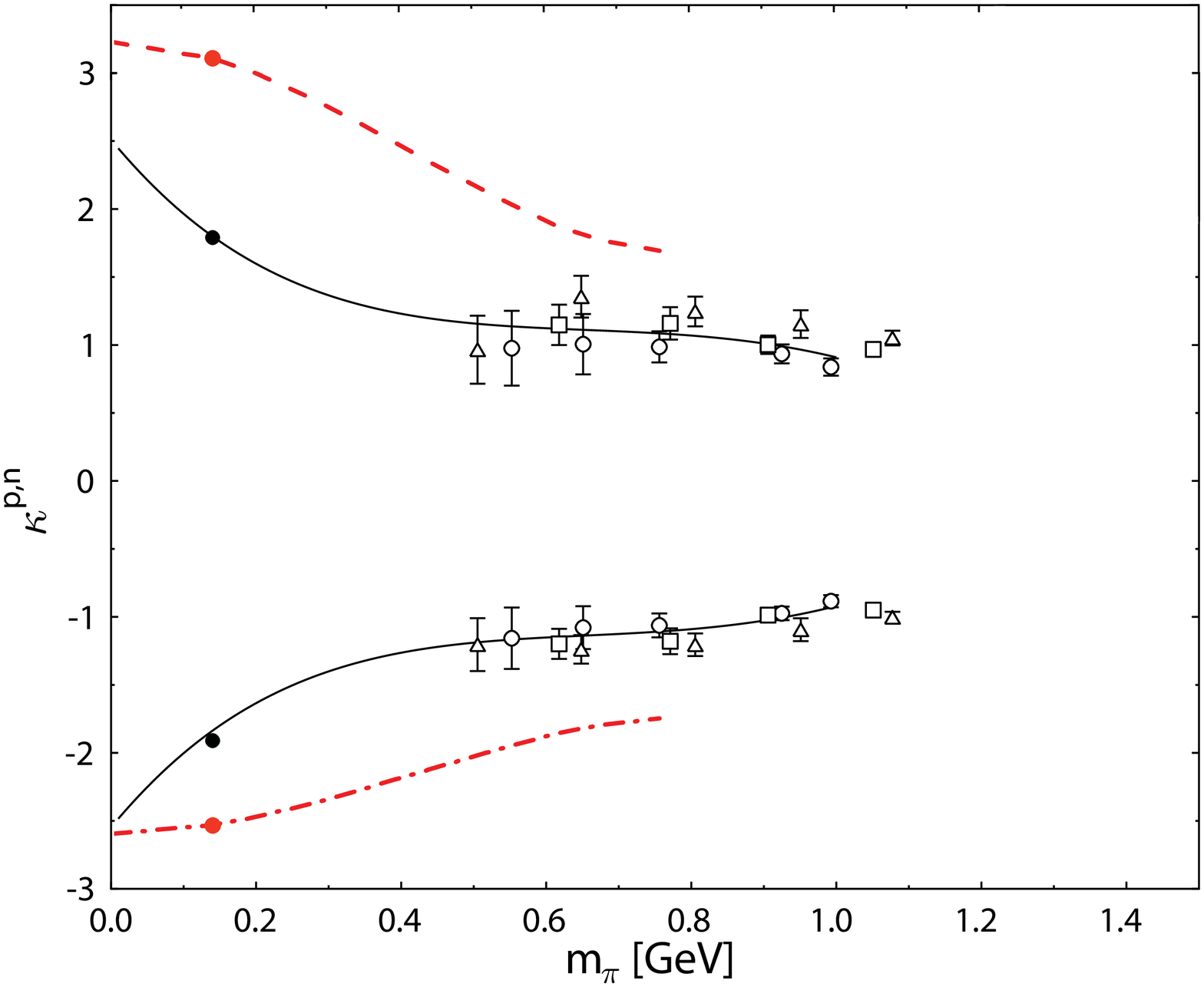}
\label{kappa}
\caption{The results for the anomalous magnetic moments (dashed lines)
as a function of the pion mass
in comparison to the experimental value and corresponding 
results from lattice data and their chiral extrapolations (fully drawn lines)
\cite{Gockeler:2003ay}.}
\end{center}
\end{figure}

The absolute value of the magnetic moments is also too large. In addition to look at
the isoscalar and isovector values one can compare directly to an estimate
without pions \cite{Hemmert:2002uh}, see table~III,  which is quite favourable
for the interpretation of having mostly missed the pion cloud. Plotting the
anomalous magnetic moments against the pion mass and comparing to
corresponding  results from lattice data and their chiral extrapolation
\cite{Gockeler:2003ay} provides further evidence for this interpretation.

\subsection{Summary of Part III}

The nucleons' mass, electromagnetic 
form factors, charge radii and magnetic moments
have been calculated  in a Poin\-car{\'e} covariant Faddeev approach

\begin{itemize}
\item[$\blacktriangleright$]
employing dressed quarks and "diquarks",
obtained from the quark DSE and diquark
  BSE (The corresponding propagators
  agree with lattice data where available.)  
\item[$\blacktriangleright$]
as well as using dressed quark- and diquark-photon vertex functions.
\item[$\blacktriangleright$]
The results describe the nucleons' quark core without pion contributions
and are obtained
\item[$\blacktriangleright$]
{\bf without model parameters},
  the only input scale is $\Lambda_{\tt QCD}$    
  {($\Lambda_{\tt MOM}$ into DSEs, $\Lambda_{\tt L}$ or
  $r_0$ in lattice calculations)}.
\end{itemize}

\section{Outlook}

In these lectures I have attempted to demonstrate the progress made within
functional approaches to strong continuum QCD and its application to hadronic
physics. Of course, further improvements are possible and highly desirable at
every step of these investigations. 

Within Landau gauge a consistent qualitative picture for all QCD propagators
and vertex functions has emerged. First results within Coulomb gauge, see also
the lectures by Dan Zwanziger, allow for a first more general interpretation of
the results. The importance of the Gribov horizon in both gauges can possibly
give a hint about the nature of confining field configurations.

Besides the fundamental issues of confinement and dynamical generation of
masses QCD Green functions can serve as a first-principles input into a 
description of hadronic properties. The ideal calculations of this kind have the
minimal parameter set of QCD and make real predictions. The results obtained so
far allow for the optimistic view that a coherent description of hadronic
states and processes based on the dynamics of confined quarks
and Yang-Mills fields can be realized in the near future. Hereby different
methods based on the QCD Green functions will yield complementary information.
Lattice calculations provide evidence that truncations of the infinite
hierarchy of Dyson--Schwinger and/or Renormalization Group equations have been
done reasonably. On the other hand, results based on the full set of these
equations as well as on truncated equations help to understand the lattice
data.

To explain confinement is one of the truly fundamental challenges to
contemporary physics. Understanding confinement will allow us to link the
microscopic degrees of freedom of QCD, the quarks and gluons, to the measurable
strong interactions in hadronic and nuclear physics. As long as the confinement
phenomenon stays mysterious the standard model of particle physics lacks an
important part, something essential in the fundamental laws of nature is then
still unknown to us. Although {\em Infrared QCD} has posed us extremely
challenging problems it is worth pursuing its research.

\acknowledgments
I cordially thank  Attilio Cucchieri, Tereza Mendes and Silvio 
P.~Sorella for inviting me to  this extraordinarily interesting conference {\em
Infrared QCD in Rio} and for their  hospitality.

The work reported was done in collaboration with a number of colleagues and
students. I am deeply thankful to all of them. I am especially grateful to
Lorenz von Smekal and Christian Fischer for sharing their insights over many
years with me and for the uncounted number of valuable discussions with them. 
I am greatly indebted to Will Detmold, Felipe Llanes-Estrada, Andreas
Krassnigg, Axel Maas, Pieter Maris, Craig Roberts, Sebastian Schmidt and Pete
Watson for their contributions  to common research which deepened my
understanding of Infared QCD and hadron physics.

Furthermore, I wish to thank Jeff Greensite, {\v S}tefan Olejn\'{\i}k and Dan
Zwanziger for many enlightening discussions about   confinement, and Peter
Tandy for helpful discussions about hadron physics.

This work is supported in part by the Deutsche Forschungsgemeinschaft
under Grant No.\ Al279/5-1  and the Austrian Science Fund FWF under Grant No.\
W1203 (Doctoral Program ``Hadrons in vacuum, nuclei and stars'').


\begin{thebibliography}{99}

\bibitem{Alkofer:2000wg}
  R.~Alkofer and L.~von Smekal,
  Phys.\ Rept.\  {\bf 353}, 281 (2001)
  [arXiv:hep-ph/0007355].

\bibitem{Fischer:2006ub}
  C.~S. Fischer, J. Phys. G: Nucl. Part. Phys. {\bf 32}, R253 (2006)
  [arXiv:hep-ph/0605173].

\bibitem{Maris:2003vk}
  P.~Maris and C.~D.~Roberts,
  Int.\ J.\ Mod.\ Phys.\ E {\bf 12}, 297 (2003)
  [arXiv:nucl-th/0301049].
  
\bibitem{Alkofer:2006fu}
  R.~Alkofer and J.~Greensite, arXiv:hep-ph/0610365.

\bibitem{Gross:1974jv}
  D.~J.~Gross and A.~Neveu,
  Phys.\ Rev.\ D {\bf 10}, 3235 (1974).
   
\bibitem{Bleuler50}
 S.~N.~Gupta, Proc.~Roy.~Soc. A{\bf 63}, 681 (1950); \newline
 K.~Bleuler, Helv.~Phys.~Acta {\bf 23}, 567 (1950).
\bibitem{Faddeev67}
 L.~D.~Faddeev and V.~N.~Popov, Phys.~Lett. B {\bf 25}, 29 (1967).
\bibitem{Becchi:1976nq}
 C.~Becchi, A.~Rouet, and R.~Stora,
 Annals Phys. {\bf 98}, 287 (1976); \newline
 {\it see also:} I. V. Tyutin, Lebedev preprint FIAN No. 39 (1975).
\bibitem{Kugo:1979gm}
 T.~Kugo and I.~Ojima, Prog.~Theor.~Phys.~Suppl. {\bf 66}, 1 (1979).
\bibitem{Nakanishi:qm}
 N.~Nakanishi and I.~Ojima,
World Sci.\ Lect.\ Notes Phys.\  {\bf 27}, 1 (1990).

\bibitem{Gribov:1978wm}
V.~N.~Gribov,
Nucl.\ Phys.\ B {\bf 139}, 1 (1978).

\bibitem{Dan} 
  see also: D.~Zwanziger, 
  Nucl.\ Phys.\ B {\bf 378}, 525 (1992).
  
\bibitem{vanBaal:1997gu}
  P.~van Baal,
  arXiv:hep-th/9711070.

\bibitem{Williams:2002dw}
  A.~G.~Williams,
  Nucl.\ Phys.\ Proc.\ Suppl.\  {\bf 109A}, 141 (2002)
  [arXiv:hep-lat/0202010].
  
\bibitem{Cucchieri:1997dx}
A.~Cucchieri, Nucl. Phys. B{\bf 508}, 353 (1997) 
[arXiv: hep-lat/9705005].

\bibitem{Bonnet:2001uh}
F.~D.~Bonnet {\it et al.},
Phys.\ Rev.\ D {\bf 64}, 034501 (2001) [arXiv:hep-lat/0101013].


\bibitem{Alexandrou:2002gs}
C.~Alexandrou, P.~De~Forcrand, and E.~Follana, { Phys. Rev.} D {\bf 65}, 117502 
(2002) [arXiv:hep-lat/0203006].


\bibitem{Silva:2004bv}
P.~J. Silva and O.~Oliveira, {Nucl. Phys.} B {\bf 690}, 177 (2004)
[arXiv:hep-lat/0403026].

\bibitem{Furui:2004cx}
S.~Furui and H.~Nakajima, {Phys. Rev.} D {\bf 70}, 094504 (2004).


\bibitem{Bowman:2004jm}
P.~O.~Bowman {\it et al.},
 and A.~G.~Williams,
Phys.\ Rev.\ D {\bf 70}, 034509 (2004)
[arXiv: hep-lat/0402032].


\bibitem{Sternbeck:2005tk}
A.~Sternbeck{\it et al.},
  Phys. Rev. D {\bf 72}, 014507 (2005) 
  [arXiv: hep-lat/0506007].

\bibitem{Zwanziger:2003cf}
  D.~Zwanziger, Phys. Rev. D {\bf 69}, 016002 (2004)
  [arXiv:hep-ph/0303028].

\bibitem{Zwanziger:1993qr}
D.~Zwanziger,
Nucl.\ Phys.\ B {\bf 364},  127 (1991); Nucl.\ Phys.\ B {\bf 399}, 477 (1993);
Nucl.\ Phys.\ B {\bf 412},  657 (1994).


\bibitem{Hata:1992dn}
  H.~Hata and I.~Niigata,
  Nucl.\ Phys.\ B {\bf 389}, 133 (1993)
  [arXiv: hep-ph/9207260].

\bibitem{Mendes:2006kc}
  T.~Mendes, A.~Cucchieri, and A. Mihara,
  arXiv:hep-lat/0611002.

\bibitem{Kugo:1995km}
T.~Kugo,
arXiv:hep-th/9511033.


\bibitem{Oehme:1980ai}
R.~Oehme and W.~Zimmermann,
Phys.\ Rev.\ D {\bf  21},  471 (1980).



\bibitem{Haag:1996}
R.~Haag,  {\em Local Quantum Physics},
Springer, 1996.

\bibitem{Watson:2001yv}
  P.~Watson and R.~Alkofer,
  Phys.\ Rev.\ Lett.\  {\bf 86}, 5239 (2001)
  [arXiv:hep-ph/0102332].
\bibitem{Lerche:2002ep}
C.~Lerche and L.~von Smekal,
Phys.\ Rev.\ D {\bf 65}, 125006 (2002)
[arXiv:hep-ph/0202194].
\bibitem{Zwanziger:2001kw}
D.~Zwanziger,
Phys.\ Rev.\ D {\bf 65},  094039 (2002)
[arXiv:hep-th/0109224].

\bibitem{Taylor:1971ff}
J.~C. Taylor, {Nucl. Phys.} B {\bf 33}, 436 (1971).

\bibitem{Cucchieri:2004sq}
A.~Cucchieri, T.~Mendes, and A.~Mihara,  {JHEP} {\bf 12}, 012 (2004)
[arXiv:hep-lat/0408034].

\bibitem{Schleifenbaum:2004id}
W.~Schleifenbaum {\it et al.},
{Phys. Rev.} D {\bf   72}, 014017 (2005)
[arXiv:hep-ph/0411052].

\bibitem{Sternbeck:2005qj}
A.~Sternbeck {\it et al.},
{PoS}   {\bf LAT2005}, 333 (2005)
[arXiv:hep-lat/0509090].

\bibitem{Alkofer:2004it}
  R.~Alkofer, C.~S.~Fischer, and F.~J.~Llanes-Estrada,
  Phys.\ Lett.\ B {\bf 611}, 279 (2005)
  [arXiv:hep-th/0412330].


\bibitem{Fischer:2006vf}
C.~S. Fischer and J.~M. Pawlowski, Phys. Rev. D {\bf 75}, 025012 (2007)
[arXiv:hep-th/0609009].

\bibitem{vonSmekal:1997is}
L.~von Smekal, R.~Alkofer, and A.~Hauck,
Phys.\ Rev.\ Lett.\  {\bf 79}, 3591 (1997) 
[arXiv:hep-ph/9705242];
L.~von Smekal, A. Hauck, and R.~Alkofer,
Annals Phys.\  {\bf 267}, 1 (1998)
[arXiv:hep-ph/9707327];
A.~Hauck, L.~von Smekal, and R.~Alkofer,
Comput.\ Phys.\ Commun.\  {\bf 112},  166 (1998)
[arXiv:hep-ph/9804376].

\bibitem{Fischer:2002hn}
C.~S.~Fischer  and R.~Alkofer,
Phys. Lett. B {\bf 536}, 177 (2002)
[arXiv:hep-ph/0202202];
C.~S.~Fischer, R.~Alkofer, and H.~Reinhardt,
Phys.\ Rev.\ D {\bf 65}, 094008 {2002} [arXiv:hep-ph/0202195];
R.~Alkofer, C.~S.~Fischer, and L.~von Smekal,
Acta Phys.\ Slov.\  {\bf 52}, 191 (2002) [arXiv:hep-ph/0205125].



\bibitem{Fischer:2003rp}
  C.~S.~Fischer and R.~Alkofer,
  Phys.\ Rev.\ D {\bf 67}, 094020 (2003)
  [arXiv:hep-ph/0301094].


\bibitem{Pawlowski:2003hq}
J.~M.~Pawlowski {\it et al.},
Phys.\ Rev.\ Lett.\  {\bf 93}, 152002 (2004)
[arXiv:hep-th/0312324].

\bibitem{Fischer:2004uk}
C.~S.~Fischer and H.~Gies,
JHEP {\bf 0410}, 048 (2004)
[arXiv:hep-ph/0408089].

\bibitem{Oliveira:2006zg}
  O.~Oliveira and P.~J.~Silva,
  these proceedings [arXiv:hep-lat/0609036].

\bibitem{MMP}
M.\ M\"uller-Preussker, these proceedings; 
  A.~Sternbeck, PhD Thesis, Humboldt Univ.\ Berlin, 2006
  [arXiv:hep-lat/0609016].
  
\bibitem{Maas:2006qw}
  A.~Maas, A.~Cucchieri, and T.~Mendes,
  these proceedings [arXiv:hep-lat/0610006].
   
\bibitem{Fischer:2005ui}
  C.~S.~Fischer, B.~Gruter, and R.~Alkofer,
  Annals Phys.\  {\bf 321}, 1918 (2006)
  [arXiv:hep-ph/0506053].
  
  
\bibitem{Fischer:2006}
C.~S.~Fischer {\it et al.}, arXiv:hep-ph/0701050.


\bibitem{Zhang:2003fa}
J.B.~Zhang {\it et al.},
  Phys.\ Rev.\ D {\bf 70}, 034505 (2004)
  [arXiv:hep-lat/0301018].
 
\bibitem{Bowman:2002bm}
  P.~O.~Bowman, U.~M.~Heller, and A.~G.~Williams,
  Phys.\ Rev.\ D {\bf 66}, 014505 (2002)
  [arXiv:hep-lat/0203001].

\bibitem{Fischer:2004wf}
  C.~S.~Fischer and R.~Alkofer,
  AIP Conf.\ Proc.\  {\bf 756}, 275 (2005)
  [arXiv:hep-ph/0411347].


\bibitem{Alkofer:2003jj}
R.~Alkofer {\it et al.},
  Phys.\ Rev.\ D {\bf 70}, 014014 (2004)
  [arXiv:hep-ph/0309077];
  Nucl.\ Phys.\ Proc.\ Suppl.\  {\bf 141}, 122 (2005).

\bibitem{Shirkov:1997wi}
D.~V.~Shirkov and I.~L.~Solovtsov,
Phys.\ Rev.\ Lett.\  {\bf 79}, 1209 (1997).


\bibitem{Maas:2005hs}
  A.~Maas, J.~Wambach, and R.~Alkofer,
  Eur.\ Phys.\ J.\ C{\bf 42}, 93 (2005)
  [arXiv:hep-ph/0504019];
A.~Maas {\it et al.},
  Eur.\ Phys.\ J.\ C{\bf 37}, 335 (2004)
  [arXiv:hep-ph/0408074];

\bibitem{Cucchieri:2003di}
  A.~Cucchieri, T.~Mendes, and A.~R.~Taurines,
  Phys.\ Rev.\ D {\bf 67}, 091502 (2003)
  [arXiv:hep-lat/0302022].

\bibitem{Alkofer:2006gz}
  R.~Alkofer, C.~S.~Fischer, and F.~J.~Llanes-Estrada,
  arXiv:hep-ph/0607293.

\bibitem{Alkofer:2005ug}
R.~Alkofer {\it et al.},
  Phys.\ Rev.\ Lett.\  {\bf 96}, 022001 (2006)
  [arXiv: hep-ph/0510028].

\bibitem{Wilson:1974sk}
  K.~G.~Wilson,
  Phys.\ Rev.\ D {\bf 10}, 2445 (1974).

\bibitem{Zwanziger:2002sh}
  D.~Zwanziger,
  Phys.\ Rev.\ Lett.\  {\bf 90}, 102001 (2003).

\bibitem{Greensite:2004ke}
  J.~Greensite, S.~Olejnik, and D.~Zwanziger,
  Phys.\ Rev.\ D {\bf 69}, 074506 (2004).

\bibitem{Nakamura:2005ux}
A.~Nakamura and T.~Saito,
Prog.\ Theor.\ Phys.\ {\bf 115}, 189 (2006) [arXiv:hep-lat/0512042].

\bibitem{Szczepaniak:2003ve}
  A.~P.~Szczepaniak,
  Phys.\ Rev.\ D {\bf 69}, 074031 (2004)
  [arXiv:hep-ph/0306030].
\bibitem{Szczepaniak:2002wk}
  A.~P.~Szczepaniak and E.~S.~Swanson,
  AIP Conf.\ Proc.\  {\bf 549}, 330 (2002).
\bibitem{Szczepaniak:2001rg}
  A.~P.~Szczepaniak and E.~S.~Swanson,
  Phys.\ Rev.\ D {\bf 65}, 025012 (2002)
  [arXiv:hep-ph/0107078].
\bibitem{Zwanziger:2003de}
  D.~Zwanziger,
  Phys.\ Rev.\ D {\bf 70}, 094034 (2004)
  [arXiv:hep-ph/0312254].
  
\bibitem{Feuchter:2004mk}
  C.~Feuchter and H.~Reinhardt,
  Phys.\ Rev.\ D {\bf 70}, 105021 (2004)
  [arXiv:hep-th/0408236].

\bibitem{Finger:1981gm}
  J.~R.~Finger and J.~E.~Mandula,
  Nucl.\ Phys.\ B {\bf 199}, 168 (1982).
\bibitem{Govaerts:1983ft}
J.~Govaerts {\it et al.},
  Nucl.\ Phys.\ B {\bf 237}, 59 (1984).
\bibitem{Adler:1984ri}
  S.~L.~Adler and A.~C.~Davis,
  Nucl.\ Phys.\ B {\bf 244}, 469 (1984).

\bibitem{Alkofer:1988tc}
  R.~Alkofer and P.~A.~Amundsen,
  Nucl.\ Phys.\ B {\bf 306}, 305 (1988).


\bibitem{Llanes-Estrada:2004wr}
F.~J.~Llanes-Estrada {\it et al.},
  Phys.\ Rev.\ C {\bf 70}, 035202 (2004)
  [arXiv:hep-ph/0402253].

\bibitem{Cucchieri:2000gu}
  A.~Cucchieri and D.~Zwanziger,
  Phys.\ Rev.\ D {\bf 65}, 014001 (2002)
  [arXiv:hep-lat/0008026].
\bibitem{Greensite:2003bk}
  J.~Greensite,
  Prog.\ Part.\ Nucl.\ Phys.\  {\bf 51}, 1 (2003)
  [arXiv:hep-lat/0301023].
\bibitem{Greensite:2003xf}
  J.~Greensite and S.~Olejnik,
  Phys.\ Rev.\ D {\bf 67}, 094503 (2003)
  [arXiv:hep-lat/0302018].
\bibitem{Greensite:2004bv}
  J.~Greensite, S.~Olejnik, and D.~Zwanziger,
  arXiv:hep-lat/0410028.
  
\bibitem{Langfeld:1989en}
  K.~Langfeld, R.~Alkofer, and P.~A.~Amundsen,
  Z.\ Phys.\ C {\bf 42}, 159 (1989).



\bibitem{Eichmann:2006}
G.\ Eichmann,
Diploma Thesis, Graz University 2006, Advisors: R.\ Alkofer
and A.\ Krassnigg  
[http://physik.uni-graz.at/ itp/DD/eichmann/diplomarbeit.pdf].


\bibitem{Alkofer:2005jh}
  R.~Alkofer and M.~Oettel,
  talk given at the Schladming Winter School 2005 
  [arXiv:nucl-th/0507003].

\bibitem{Oettel:1998bk}
M.~Oettel  {\it et al.},
  Phys.\ Rev.\ C {\bf 58}, 2459 (1998)
  [arXiv:nucl-th/9805054];
 G.~Hellstern  {\it et al.},
  Nucl.\ Phys.\ A {\bf 627}, 679 (1997)
  [arXiv:hep-ph/9705267].

\bibitem{Oettel:1999gc}
  M.~Oettel, M.~Pichowsky, and L.~von Smekal,
  Eur.\ Phys.\ J.\ A {\bf 8}, 251 (2000)
  [arXiv:nucl-th/9909082].


\bibitem{Oettel:2000ig}
  M.~Oettel, PhD thesis, T\"ubingen University 2000, Advisor: R.\ Alkofer
  [arXiv:nucl-th/0012067].


\bibitem{Eichmann:2007} R.~Alkofer, G.~Eichmann, and A.~Krassnigg,
in preparation.

\bibitem{Oettel:2001kd}
  M.~Oettel, L.~von Smekal, and R.~Alkofer,
  Comput.\ Phys.\ Commun.\  {\bf 144}, 63 (2002)
  [arXiv:hep-ph/0109285].


\bibitem{Gockeler:2003ay}
  M.~Gockeler {\it et al.}  
                  [QCDSF Collaboration],
  Phys.\ Rev.\ D {\bf 71}, 034508 (2005)
  [arXiv:hep-lat/0303019].
  

\bibitem{Hecht:2002ej}
M.~B.~Hecht {\it et al.},
  Phys.\ Rev.\ C {\bf 65}, 055204 (2002)
  [arXiv:nucl-th/0201084].

\bibitem{Oettel:2002cw}
  M.~Oettel and A.~W.~Thomas,
  Phys.\ Rev.\ C {\bf 66}, 065207 (2002)
  [arXiv:nucl-th/0203073].

\bibitem{Kvinikhidze:1999xn}
A.~N.~Kvinikhidze and B.~Blankleider,
Phys.\ Rev.\ C {\bf 60}, 044003 (1999)
[nucl-th/9901001].

  
\bibitem{Oettel:2000jj}
  M.~Oettel, R.~Alkofer, and L.~von Smekal,
  Eur.\ Phys.\ J.\ A {\bf 8}, 553 (2000)
  [arXiv:nucl-th/0006082].

\bibitem{Ball:1980ay}
J.~S.~Ball and T.~Chiu,
Phys.\ Rev.\ D {\bf 22}, 2542 (1980).

\bibitem{Curtis:1990zs}
D.~C.~Curtis and M.~R.~Pennington,
Phys.\ Rev.\ D {\bf 42}, 4165 (1990).

\bibitem{Maris:2000bh}
P.~Maris and P.~C.~Tandy,
Phys.\ Rev.\ C {\bf 61}, 045202 (2000)
[nucl-th/9910033].

\bibitem{Weiss:1993kv}
C.~Weiss  {\it et al.},
  Phys.\ Lett.\ B {\bf 312}, 6 (1993)
  [arXiv:hep-ph/9305215].

\bibitem{Ahlig:2000qu}
S.~Ahlig {\it et al.},
  Phys.\ Rev.\ D {\bf 64}, 014004 (2001)
  [arXiv:hep-ph/0012282].
  
\bibitem{Oettel:2002wf}
  M.~Oettel and R.~Alkofer,
  Eur.\ Phys.\ J.\ A {\bf 16}, 95 (2003)
  [arXiv:hep-ph/0204178].

    
\bibitem{Alkofer:2004yf}
R.~Alkofer {\it et al.},
  Few Body Syst.\  {\bf 37}, 1 (2005)
  [arXiv:nucl-th/0412046];
A.~H\"oll {\it et al.},
  Nucl.\ Phys.\ A {\bf 755}, 298 (2005)
  [arXiv:nucl-th/0501033].

\bibitem{Hemmert:2002uh}
  T.~R.~Hemmert and W.~Weise,
  Eur.\ Phys.\ J.\ A {\bf 15}, 487 (2002)
  [arXiv:hep-lat/0204005].

\end{thebibliography}
\end{document}